\documentclass[%
 reprint,
superscriptaddress,
nofootinbib,
 amsmath,amssymb,
 aps,
]{revtex4-2}

\usepackage{graphicx}
\usepackage{dcolumn}
\usepackage{dsfont}
\usepackage{bbold}
\usepackage{xcolor}
\usepackage{braket}
\usepackage{multirow}
\usepackage[colorlinks=true, linkcolor=blue, citecolor=blue, urlcolor=blue]{hyperref}
\usepackage{todonotes} 
\usepackage[normalem]{ulem} 

\begin{document}

\title{Path integral approach to the truncated Wigner approximation\\ of driven-dissipative spins}%

	\author{Viktoria Noel}
    \email[]{viktoria.noel@uni-tuebingen.de}
    \affiliation{Institute for Theoretical Physics, and Center for Integrated Quantum Science and Technology, Universität Tübingen, Auf der Morgenstelle 14, 72076 Tübingen, Germany}
    \author{Michael Fleischhauer}
    \affiliation{Department of Physics and Research Center OPTIMAS, RPTU University Kaiserslautern-Landau, D-67663 Kaiserslautern, Germany}
	\author{Igor Lesanovsky}
	\affiliation{Institute for Theoretical Physics, and Center for Integrated Quantum Science and Technology, Universität Tübingen, Auf der Morgenstelle 14, 72076 Tübingen, Germany}
    \affiliation{School of Physics and Astronomy and Centre for the Mathematics and Theoretical Physics of Quantum Non-Equilibrium Systems, University of Nottingham, Nottingham, NG7 2RD, United Kingdom}

\date{\today}

\begin{abstract}

Phase-space approaches such as the truncated Wigner approximation (TWA) provide an efficient semiclassical framework for performing approximate simulations of the dynamics of open quantum many-body systems outside the reach of exact numerical methods but beyond the mean-field level.
For bosonic systems, TWA is known to be equivalent to a Keldysh path-integral formulation truncated at second order in the so-called quantum fluctuations.
This semiclassical approach provides an alternative transparent route towards approximate stochastic equations of motion, which can be efficiently solved.
Here we establish the corresponding path-integral formulation for interacting open spin-$1/2$ systems using the continuous $\mathrm{SU}(2)$ phase space. 
We show, in particular, that a consistent treatment of dissipation requires correctly mapping operator products onto the curved spin phase space, leading to stochastic equations that coincide with those obtained from the continuous TWA formulation and thus reproduce the exact dynamics of a single dissipative spin.  
Our results provide a unified field-theoretic foundation for the TWA to dissipative spin dynamics and offer a systematic starting point for extensions beyond the semiclassical approximation.

\end{abstract}

\maketitle


\section{Introduction}

Understanding the nonequilibrium dynamics of interacting quantum systems coupled to an environment is a central challenge in modern many-body physics \cite{Weimer_2021, Fazio_2025, Sieberer_2025, Harrington_2022, Diehl_2008, Mi_2024}. 
Advances in experimental platforms ranging from ultracold atoms \cite{gross2017quantum, bloch2012quantum, Lewenstein_2007} to trapped ions \cite{blatt2012quantum, schneider2012experimental, Monroe_2021}, superconducting circuits \cite{Houck_2012, Blais_2021, Kjaergaard_2020} and cavity QED systems \cite{Walther_2006, Mivehvar_2021, Ritsch_2013} have made it possible to engineer driven-dissipative quantum matter with an unprecedented degree of control. 
These systems exhibit a rich variety of collective phenomena arising from the interplay of coherent interactions, external driving and dissipation, such as dissipative phase transitions \cite{Kessler_2012, Minganti_2018}, bistability \cite{Drummond_1980, Carr_2013}, synchronisation \cite{Lee_2013, Walter_2014}, super- and subradiance \cite{Dicke_1954, Gross_1982, Scully_2015, Guerin_2016}, and time-crystalline order \cite{Iemini_2018}.
Their dynamics is commonly described by Lindblad master equations, which provide an effective Markovian description after eliminating the environmental degrees of freedom.
However, Lindblad equations of interacting many-body systems remain notoriously difficult to solve. 
The dimension of the density matrix grows exponentially with system size, severely limiting exact numerical approaches. 
Approximate techniques, such as tensor-network methods \cite{Schollwoeck:2010uqf,Verstraete_2004, Zwolak_2004}, cumulant expansions \cite{Kubo:1962gce,Plankensteiner:2021wst} and diagrammatic approaches \cite{sieberer2016keldysh, kamenev2023field}, can substantially extend the accessible regimes, but often require problem-specific adaptations and become challenging in strongly correlated or higher-dimensional settings, and are numerically costly. 
Developing simple and efficient approximations that retain the essential quantum effects while remaining computationally tractable therefore remains an important goal.

Phase-space methods provide an appealing route towards this objective.
In particular, the truncated Wigner approximation (TWA) has proven highly successful for isolated bosonic systems \cite{Sinatra_2002, Blakie_2008, polkovnikov2010phase}, where quantum dynamics is approximated by an ensemble of classical trajectories with quantum fluctuations encoded in the initial conditions. 
The approach has also been applied to spin systems through continuous phase-space formulations, e.g., via the Schwinger-boson representation of spins \cite{polkovnikov2010phase}, together with closely related semiclassical treatments based on spin-coherent-state path integrals \cite{solari1987semiclassical, kochetov19952}. These provide descriptions that are controlled in the large-spin limit \cite{polkovnikov2010phase}.
Formulations based on discrete phase spaces have extended the approach to interacting spin-$1/2$ models without requiring a large-spin expansion, and have found numerous applications in nonequilibrium quantum dynamics \cite{Schachenmayer_2015}.
Such interacting spin-$1/2$ models are directly realised in Rydberg atom arrays, where laser excitation to a Rydberg state realises a two-level spin coupled through strong, long-range interactions \cite{Weimer_2010, Saffman_2010, Browaeys_2020}. 

However, any real-world system inevitably couples to its environment.
In Rydberg arrays, for instance, radiative decay renders the dynamics intrinsically driven-dissipative, making it essential to extend phase-space methods beyond the isolated systems for which they were originally devised.
More recently, several works have generalised the TWA to open quantum systems and demonstrated that stochastic phase-space approaches can provide accurate descriptions of dissipative many-body dynamics over a broad range of parameters \cite{Qu_2019, Liu_2020, Huber_2021, Huber_2022, mink2022hybrid, Singh_2022, Mink_2023, Tebbenjohanns_2024, Zhang:2025kjj, Ruks:2026ndx, Mondal:2026meg, xiang2025quantumpredatorpreycyclesdissipative, hosseinabadi2025user, noel2026quantum, hartmann2026truncatedwignerapproximationspins}.

These developments have established dissipative TWA as a powerful numerical tool, and recent work has derived dissipative spin TWA from Keldysh field theory \cite{hosseinabadi2025user}, adapting the bosonic open system path integrals of \cite{sieberer2016keldysh}.
In any such formulation, however, mapping operator products from Hilbert space onto phase space yields not the ordinary multiplicative product of the operators' phase-space symbols but a so-called star product, which supplements it with corrections that encode the noncommutativity of the operators \cite{moyal1949quantum, varilly1989moyal}.
For bosons, whose phase space is flat, these corrections are organised in powers of $\hbar$, which in a dimensionless formulation corresponds to inverse powers of the occupation of modes.
Truncating these at leading order is precisely the controlled first step of the semiclassical expansion \cite{polkovnikov2010phase}.
For a spin of length $S$, whose phase space is curved, the corresponding dimensionless expansion parameter is $1/S$ \cite{polkovnikov2010phase, klimov2002moyal}, and in the limit $S\to\infty$, where the phase space becomes effectively flat, the star product reduces to the ordinary product of the symbols, in analogy with the semiclassical limit for bosons \cite{klimov2002moyal}. 

However, for spin-$1/2$, the corrections generated by the star product on the $\mathrm{SU}(2)$ phase space are of the same order as the ordinary product itself, so there is \textit{a priori} no small parameter that controls the error incurred by neglecting them.
Existing field-theoretic frameworks of dissipative spin systems do not consider these corrections \cite{hosseinabadi2025user}, which can have a dramatic impact on the accuracy of TWA simulations.
Failing to treat the star product appropriately can yield an inconsistent semiclassical limit, which is already apparent at the level of a single spin.
This has been overcome by the continuous truncated Wigner approach, which reproduces the exact dynamics of single-spin decay \cite{mink2022hybrid}.
A field-theoretic formulation that systematically incorporates the $\mathrm{SU}(2)$ star product is therefore desirable, both to make the underlying approximations transparent and to provide a route towards higher-order corrections.

In this work, we develop a path-integral formulation of the TWA for driven-dissipative many-body spin-$1/2$ systems based on the continuous $\mathrm{SU}(2)$ phase space. 
A central result of our analysis is that the $\mathrm{SU}(2)$ star product generates additional $\mathcal O(1)$ contributions to the dissipative parts of the resulting equations of motion, with no $1/S$ suppression available at $S=1/2$.
These additional terms are not captured when phase-space symbols are multiplied as ordinary functions and are essential for obtaining a consistent semiclassical description.
We further demonstrate that the corrected stochastic equations reproduce the exact dynamics of a single dissipative spin, in agreement with the continuous truncated Wigner approach \cite{mink2022hybrid,hartmann2026truncatedwignerapproximationspins}.

This paper is organised as follows:
Section~\ref{sec:pathintegral} introduces the $\mathrm{SU}(2)$ phase space of a two-state system, on which we then construct the path integral for the Lindblad evolution.
In Sec.~\ref{sec:singlespin} we focus on a single driven-decaying spin, obtain the semiclassical action, and show that the stochastic equations of motion derived from this reproduce the exact Lindblad dynamics of the spin expectation values, whereas replacing the star product by the ordinary product does not.
Section~\ref{sec:manybody} extends the construction to interacting spins of Rydberg arrays, where the truncation of the interaction terms constitutes the only approximation.
In Sec.~\ref{sec:results} we benchmark the resulting equations against exact master-equation solutions, for the single spin and for a driven-dissipative Rydberg chain, isolating the role of the star-product corrections in the dissipative sector.
Section~\ref{sec:summary} presents a summary and an outlook.

\section{The $\mathrm{SU(2)}$ path integral approach to the TWA}
\label{sec:pathintegral}

\begin{figure}[t]
\includegraphics[width=\linewidth]{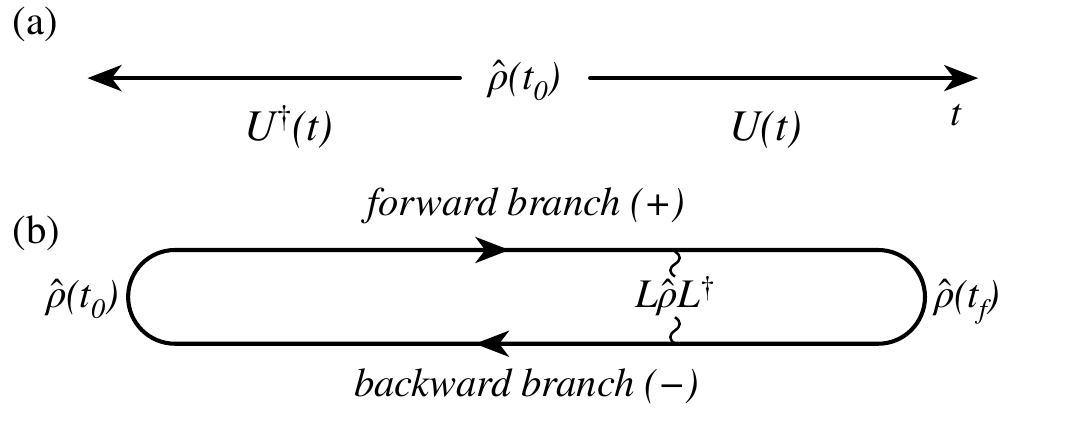}
\caption{\textbf{The Keldysh contour for Lindblad dynamics.} (a): Evolution of the density matrix,
$\hat\rho(t)=\hat U(t)\,\hat\rho(t_0)\,\hat U^\dagger(t)$: the initial state is propagated by $\hat U(t)$ on one side and by $\hat U^\dagger(t)$ on the other, so its evolution comprises two histories over the same time interval.
(b): The two histories folded into the closed time contour: the system travels along the forward branch ($+$) and returns along the backward branch ($-$). 
The initial state $\hat\rho(t_0)$ joins the two branches at the initial time, and the closure of the contour at the final time is implemented by computing the trace of $\hat\rho(t_f)$. Dissipation couples the branches: in the term $\hat L\hat\rho\hat L^\dagger$, $\hat L$ acts on the forward and $\hat L^\dagger$ on the backward branch at the same instant.}
\label{fig:contour}
\end{figure}

In this section, we construct a path integral for driven-dissipative spin-$1/2$ systems directly on the continuous $\mathrm{SU}(2)$ phase space. 
Our starting point is the Lindblad quantum master equation, which describes the time evolution of the reduced system density matrix $\hat \rho$ and reads
\begin{equation}
  \partial_t\hat\rho = \mathcal{L}\hat\rho
  =-i[\hat H,\hat\rho]
  \;+\; \sum_{\alpha}\!\left(
      \hat L_\alpha\hat\rho \hat L_\alpha^\dagger
      - \tfrac12\{\hat L_\alpha^\dagger \hat L_\alpha,\hat\rho\}\right).
  \label{eq:lindblad}
\end{equation}
The commutator term, familiar from the von Neumann equation, generates the coherent dynamics of the system Hamiltonian $\hat H$.
The second part, the dissipator, describes the coupling of the system to its environment, encoded in a set of jump operators $\hat L_\alpha$. 

The expectation value of an operator $\hat O$ is defined as $\langle \hat O\rangle(t)=\operatorname{tr}[\hat O\,\hat \rho(t)]$, and already for unitary dynamics, $\hat \rho(t)=\hat U(t)\hat \rho(0) \hat U^\dagger(t)$, this puts one evolution operator to the \emph{left} of the initial state and its conjugate to the \emph{right}, with the trace tying the two free ends together. 
Written as a path integral, the left operator $\hat U$ sums over histories running forward from $0$ to $t$ and the right operator $\hat U^\dagger$ sums over histories running backward from $t$ to $0$; the trace identifies their endpoints. 
A natural description for this ordering is offered by the Keldysh contour, since the system travels along a forward branch and back along a backward branch, and the two are joined into a single closed time contour.
This is illustrated in Fig.~\ref{fig:contour}.
Dissipation does not change this picture, it only couples the two branches: in Eq.~\eqref{eq:lindblad} the term $\hat L \hat \rho \hat L^\dagger$ has $\hat L$ acting on $\hat \rho$ from the left (the forward side) and $\hat L^\dagger$ from the right (the backward side) at the same instant.
The formal path integral construction based on this is described in detail in \cite{sieberer2016keldysh} for driven-dissipative bosonic systems. 
The corresponding TWA/semiclassical approximation\footnote{For an alternative path integral approach to obtain the TWA of bosonic open quantum systems, we refer to \cite{yoneya2025path}. Instead of the bosonic coherent state path integral in \cite{sieberer2016keldysh}, \cite{yoneya2025path} formulates the path integral directly from the Wigner--Weyl transform of the Lindblad master equation, reducing the resulting Fokker--Planck equation to stochastic trajectories. This is close in spirit to the construction we develop below.} of this construction is presented in \cite{hosseinabadi2025user}, where it was also applied to spin systems.

To obtain the TWA, on the Keldysh contour, one rotates the forward and backward branches to the ``classical--quantum'' basis and expands the action in powers of the quantum fields, which has been shown to be equivalent to an expansion in $\hbar$ \cite{polkovnikov2010phase, kamenev2023field}, and in $1/S$ for a spin of length $S$ \cite{polkovnikov2010phase, solari1987semiclassical, kochetov19952}. 
The semiclassical approximation retains terms at most quadratic in the quantum fields, which reduces the path integral to classical trajectories with stochastic initial conditions and, in the presence of dissipation, stochastic noise along the trajectories. 
Applied to driven-dissipative spin systems, however, this bosonic construction \cite{hosseinabadi2025user} fails in some instances: both for a single spin-$1/2$ and in the many-body case, it lands on the wrong steady state even for a purely decaying spin, where the exact steady state coincides with the classical one, as we detail in Sec.~\ref{sec:results}.
In such a regime the semiclassical truncation would not be expected to fail, which hints that the terms discarded for spins were not, in fact, higher order.

\subsection{From Hilbert to phase space for two-state quantum systems}
The goal of any phase-space formulation of quantum mechanics is to trade operators on Hilbert space for ordinary functions on a classical phase space: observables become functions of the phase-space coordinates, the state becomes a quasiprobability distribution, and quantum expectation values become phase-space averages.
For a continuous degree of freedom, this is the familiar Wigner--Weyl correspondence for the bosonic (flat) phase space \cite{wigner1932quantum, weyl1927quantenmechanik}.
In general, such a correspondence is generated by a pair of kernels (or quantisers), which are operator-valued functions on phase space.

For a spin-$1/2$ this construction requires some care, as the phase space is compact and there is no large-$S$ classical limit to lean on.
We adopt the Stratonovich--Weyl kernel \cite{stratonovich1957distributions, varilly1989moyal}
\begin{equation}
    \hat\Delta(\Omega)=\tfrac12(\mathbb 1+\sqrt3\,\mathbf n(\Omega)\!\cdot\!\boldsymbol{\hat{\sigma}}).
    \label{eq:swkernel}
\end{equation}
Here, $\boldsymbol{\hat{\sigma}}=(\hat{\sigma}^x, \hat{\sigma}^y, \hat{\sigma}^z)$ is the Pauli vector, and 
\begin{equation}
\mathbf n=(\sin\theta\cos\phi,\sin\theta\sin\phi,\cos\theta),
\label{eq:nvector}
\end{equation}
the unit vector. 
The four-point discretisation of Eq.~\eqref{eq:swkernel} reproduces the phase-point operators of Wootters \cite{wootters1987wigner}.
The kernel \eqref{eq:swkernel} is a Hermitian, unit-trace operator attached to each point $\Omega=(\theta,\phi)$ of the Bloch sphere, which assigns to every operator $\hat{O}$ a $c$-number symbol $O(\Omega)$, and being self-dual, also reconstructs the operator from its symbol as 
\begin{equation}
O(\Omega)=\operatorname{tr}[\hat\Delta(\Omega)\hat O]; \quad \hat O=\!\int\! d\mu\,O(\Omega)\hat\Delta(\Omega)
\label{eq:symbol}
\end{equation}
with the normalised measure $ d\mu=(1/2\pi )\sin\theta\,d\theta\,d\phi$, normalised to $\int d\mu = 2$ for convenience.
Self-duality also implies that traces of operator products become ordinary phase-space averages, 
\begin{equation}
\operatorname{tr}[\hat O\hat P]=\!\int\! d\mu\,O(\Omega)P(\Omega).
\label{eq:selfdual}
\end{equation}
The density matrix $\hat\rho$ can also be written in terms of its $c$-number symbol $W$ and reconstructed from it via
\begin{equation}
W(\Omega)= \mathrm{tr}[\hat\Delta(\Omega)\hat\rho], \quad \hat\rho = \!\int d\mu\,W(\Omega)\,\hat\Delta(\Omega), 
\label{eq:density-sw}
\end{equation}
where $W$ is the Wigner function (the symbol of $\hat\rho$).
Then, expectation values simply follow from \eqref{eq:selfdual} as 
\begin{equation}
\langle \hat O \rangle = \mathrm{tr}[\hat O \hat \rho] = \int d \mu W(\Omega)O(\Omega).
\end{equation}
However, the symbol map does not preserve products: the symbol of an operator product $\hat O\hat P$ is not the ordinary product $OP$ of the individual symbols.
Evaluating \eqref{eq:symbol} on $\hat O\hat P$ instead defines a new, noncommutative multiplication rule for the symbols themselves, the $\mathrm{SU}(2)$ \emph{star} (Moyal-type) product \cite{moyal1949quantum, varilly1989moyal},
\begin{equation}
  (O\star P)(\Omega)\;\equiv\;\mathrm{tr}\bigl[\hat\Delta(\Omega)\,\hat O\hat P\bigr].
  \label{eq:star}
\end{equation}
Like its flat phase-space counterpart, the Moyal product, the star product acts as a differential operator on the symbols: generally, it can be written as the sum of the ordinary product $O(\Omega)P(\Omega)$ and correction terms that contain derivatives of $O$ and $P$ \cite{varilly1989moyal}.
Since the spin Hilbert space is finite-dimensional, these corrections terminate at finite order, which we make explicit in Sec.~\ref{sec:construction}.
Whether the star product may be replaced by the ordinary product is therefore a question about the magnitude of these derivative terms, and as mentioned in the previous section, for spin $S=1/2$, neglecting the corrections is an uncontrolled approximation.

\subsection{The construction of the path integral}
\label{sec:construction}

To construct the path integral explicitly, one starts with the partition function
\begin{equation}
Z \;=\; \mathrm{tr}\,\hat{\rho}(t_f)
\;=\;
\mathrm{tr}\Bigl[\mathcal T\exp\Bigl(\textstyle\int_{t_0}^{t_f}\!dt\,\mathcal L\Bigr)\hat{\rho}(t_0)\Bigr],
\end{equation}
and splits $[t_0,t_f]$ into $N$ slices of width $\delta t$,
\begin{equation}
\hat{\rho}(t_f) \;=\; \bigl(e^{\mathcal L\delta t}\bigr)^N\hat{\rho}(t_0),
\quad
e^{\mathcal L\delta t} = 1 + \delta t\,\mathcal L + \mathcal O(\delta t^2).
\label{eq:trotter}
\end{equation}
We now build the path integral following the state slice by slice.
The Trotter product \eqref{eq:trotter} factorises the evolution into $N$ steps of width $\delta t$, each generated by a single application of $\mathcal{L}$.
After the $n$th step, the density matrix becomes
\begin{equation}
    \hat \rho_{n+1} = (1+\delta t \mathcal{L}) \hat \rho_n +\mathcal{O}(\delta t^2).
    \label{eq:next-step-trotter}
\end{equation}
At each slice, the state is resolved on the sphere. 
Rather than expanding in $W$ with the curved measure $d\mu$, we follow Ref.~\cite{mink2022hybrid} and expand in the ``flattened'' Wigner function $\chi(\Omega)=\sin\theta\,W(\Omega)/2\pi$ with respect to the flat measure $d\Omega=d\theta\,d\phi$.\footnote{As shown in \cite{mink2022hybrid}, it is $\chi$, not $W$, whose evolution is of (generalised) Fokker--Planck type. The flattened Wigner function could be avoided by introducing a contravariant coordinate vector and a metric tensor of the spherical phase space, however both approaches yield the same differential equations.} 
Each slice is thus resolved as
\begin{equation}
\hat\rho_n=\int d\theta_n\,d\phi_n\;\chi_n(\Omega_n)\,\hat\Delta(\Omega_n),
\label{eq:flat-res}
\end{equation}
with $\chi$ defined on the strip $\theta\in[0,\pi]$, $\phi\in[0,2\pi)$. 
Evolving the system amounts to finding the step from $\chi_n$ to $\chi_{n+1}$; iterating over all slices builds the path integral.
Substituting \eqref{eq:flat-res} into \eqref{eq:next-step-trotter},
\begin{equation}
\hat\rho_{n+1}=\int d\theta_n\,d\phi_n\;\chi_n\,(1+\delta t\,\mathcal L)\,\hat\Delta_n,
\label{eq:flat-step}
\end{equation}
with $\hat\Delta_n\equiv\hat\Delta(\Omega_n)$ and $\chi_n\equiv\chi(\Omega_n)$.
The Liouvillian is linear, so it acts on the kernel $\hat\Delta_n$ and not on the scalar function $\chi_n$.
The whole construction now rests on evaluating the step $\mathcal L\hat\Delta_n$, which we show in the following.

Every term of \eqref{eq:lindblad} multiplies its argument by fixed operators, from the left or from the right.
When the argument is the kernel, the result of such a multiplication can itself be expressed through the kernel.
At every point away from the poles ($\theta=0,\pi$) the four matrices 
\begin{equation}
\mathcal B=\{\hat\Delta,\partial_\theta\hat\Delta,\partial_\phi\hat\Delta, \partial_\phi^2\hat\Delta \}
\label{eq:der-basis}
\end{equation}
span the space of $2\times2$ matrices (see App.~\ref{app:diff-id}).
We note that the choice of differential operators in Eq. \eqref{eq:der-basis} is not unique and depending on the problem other choices may be more appropriate. For a given set $\mathcal B$ there are then unique coefficient functions so that for any operator $\hat X$ 
\begin{equation}
\hat X\hat\Delta(\Omega)=\mathcal D^{L}_{X}\hat\Delta(\Omega),
\qquad
\hat\Delta(\Omega)\hat X=\mathcal D^{R}_{X}\hat\Delta(\Omega).
\label{eq:onesided}
\end{equation}
Here, $\mathcal D^{L,R}_{X}$ are differential operators of finite order in $(\theta,\phi)$ with $c$-number coefficients, and the subscript labels the operator $\hat X$ being represented, while the superscript indicates whether $\hat X$ multiplies the kernel from the left ($L$) or from the right ($R$).
These identities are the $\mathrm{SU}(2)$ star product \eqref{eq:star} in operator form.
To provide an explicit example, consider the symbol of a product $\hat X\hat P$: cycling the trace moves $\hat X$ next to the kernel, where \eqref{eq:onesided} applies, and the differential operator, having $c$-number coefficients, can be moved outside of the trace,
\begin{equation}
(X\star P)(\Omega)
=\operatorname{tr}\bigl[\hat P\,\hat\Delta(\Omega)\hat X\bigr]
=\mathcal D^{R}_{X}\operatorname{tr}\bigl[\hat P\hat\Delta(\Omega)\bigr]
=\mathcal D^{R}_{X}\,P(\Omega),
\label{eq:star-onesided}
\end{equation}
and analogously $(P\star X)(\Omega)=\mathcal D^{L}_{X}\,P(\Omega)$; note that cycling places $\hat X$ on the opposite side of the kernel, so left and right multiplication swap roles.
Since the basis \eqref{eq:der-basis} carries at most two derivatives, the operators $\mathcal D^{L,R}_{X}$ are of at most second order, and hence so is the star product itself, with all the terms contributing at $\mathcal O(1)$.
Applied term by term to the Lindbladian, namely a left operator $\mathcal D^{L}$ for every factor multiplying from the left, and a right operator $\mathcal D^{R}$ for every factor multiplying from the right, we get
\begin{equation}
\begin{aligned}
\mathcal L\hat\Delta_n
={}&\Bigl[-i\bigl(\mathcal D^{L}_{H}-\mathcal D^{R}_{H}\bigr)
+\sum_\alpha\Bigl(\mathcal D^{L}_{L_\alpha}\mathcal D^{R}_{L^\dagger_\alpha}\\
&-\tfrac12\,\mathcal D^{L}_{L^\dagger_\alpha L_\alpha}
-\tfrac12\,\mathcal D^{R}_{L^\dagger_\alpha L_\alpha}\Bigr)\Bigr]\hat\Delta_n.
\end{aligned}
\label{eq:kernel-lindblad}
\end{equation}
In the following, we will refer to the terms descending from $\hat L_\alpha\hat\rho\hat L^\dagger_\alpha$ as the jump entry of the dissipator, while the terms descending from the anticommutator, $\frac{1}{2}\left\{\hat L_\alpha^{\dagger} \hat L_\alpha, \hat \rho\right\}$, will be called ``no-jump'' terms.

The update \eqref{eq:flat-step} propagates a single function, and the resolution \eqref{eq:flat-res} uses one sphere per time slice. 
To chain the slices from the update \eqref{eq:flat-step} into a path integral, $\hat \rho_{n+1}$ is rewritten as a resolution at the new point $\Omega_{n+1}$. 
Inserting $\hat\Delta(\Omega_n)=\int d\Omega_{n+1}\,\delta(\Omega_{n+1}-\Omega_n)\,\hat\Delta(\Omega_{n+1})$ and representing the $\delta$-function as a Fourier integral\footnote{Although $\phi$ is a periodic coordinate, the Fourier representation on the line is exact here: both $\phi_n$ and $\phi_{n+1}$ lie in a single period, so $|\Delta\phi|<2\pi$ and the $\delta$-function has its only zero at $\Delta\phi=0$.}, one variable per coordinate, we get
\begin{equation}
\delta(\Omega_{n+1}-\Omega_n)=\!\int\!\frac{d\tilde\theta\,d\tilde\phi}{(2\pi)^2}\,
e^{\,i\bigl[\tilde\theta(\theta_{n+1}-\theta_n)+\tilde\phi(\phi_{n+1}-\phi_n)\bigr]}. 
\label{eq:fourierdelta}
\end{equation}
Here we introduce the pair $(\tilde\theta,\tilde\phi)$ conjugate to the elementary step.
As it will become apparent once the action is assembled, the pair $(\tilde\theta,\tilde\phi)$ can be identified \emph{a posteriori} with the response fields of the Martin--Siggia--Rose--Janssen--De Dominicis (MSRJD) representation of stochastic dynamics \cite{martin1973statistical,
janssen1976lagrangean, de1976technics}.
In Keldysh field theory this structure is associated with the semiclassical limit, as shown for bosons before: truncating the action by discarding terms cubic or higher in the quantum fields leaves an action of MSRJD form, with the quantum fields playing the role of response fields \cite{kamenev2023field, sieberer2016keldysh}. 

Under this insertion the entire $\Omega_n$-dependence of the kernel $\hat\Delta(\Omega_n)$ sits in the $\delta$-function.
The derivatives contained in the operators of \eqref{eq:kernel-lindblad} act on the exponential and simply multiply it,
\begin{equation}
\partial_{\theta_n}e^{\,i[\tilde\theta\Delta\theta+\tilde\phi\Delta\phi]}
=-i\tilde\theta\,e^{\,i[\tilde\theta\Delta\theta+\tilde\phi\Delta\phi]},
\quad
\partial_{\phi_n}\to-i\tilde\phi,
\label{eq:prepoint}
\end{equation}
with $\Delta\theta\equiv\theta_{n+1}-\theta_n$, $\Delta\phi\equiv\phi_{n+1}-\phi_n$.
Since the operators \eqref{eq:onesided} are finite order differential operators with $c$-number coefficients, this substitution turns each of them into a polynomial in $(\tilde\theta,\tilde\phi)$ whose coefficients are functions of $\Omega_n$.
Collecting the coefficient of $\hat\Delta(\Omega_{n+1})$ in
\eqref{eq:flat-step} after the delta function insertion, the slice update $\chi_{n+1}$ becomes
\begin{equation}
\chi_{n+1}=\!\int\!d\Omega_n\!\int\!\frac{d\tilde\theta\,d\tilde\phi}{(2\pi)^2}\,
e^{\,i[\tilde\theta\Delta\theta+\tilde\phi\Delta\phi]}
\bigl[1+\delta t\,\ell(\Omega_n;\tilde\theta,\tilde\phi)\bigr]\,\chi_n,
\label{eq:stepweight}
\end{equation}
with the generator symbol read off termwise from \eqref{eq:kernel-lindblad},
\begin{equation}
\begin{aligned}
\ell(\Omega;\tilde\theta,\tilde\phi)
={}&-i\bigl(H_+-H_-\bigr)
+\sum_\alpha\Bigl[
\bigl(L_\alpha\!\cdot\!L^\dagger_\alpha\bigr)_{+-}\\
&-\tfrac12\bigl(L^\dagger_\alpha L_\alpha\bigr)_+
-\tfrac12\bigl(L^\dagger_\alpha L_\alpha\bigr)_-\Bigr].
\end{aligned}
\label{eq:ell-sided}
\end{equation}
Here, $(L_\alpha \cdot L^\dagger_\alpha)_{+-}$ denotes the polynomial of the two-sided multiplication $\hat L_\alpha\hat\Delta\hat L^\dagger_\alpha$, which is the composition $\mathcal D^{L}_{L_\alpha}\mathcal D^{R}_{L^\dagger_\alpha}$ expanded in the basis \eqref{eq:der-basis} and substituted as in \eqref{eq:prepoint}.
This will be evaluated carefully for each jump operator, since it is not necessarily simply the ordinary multiplicative product of the operators.
In the above, we have also introduced for an operator $\hat X$ its forward and backward symbols $X_\pm(\Omega_n;\tilde\theta,\tilde\phi)$ as the polynomials obtained from $\mathcal D^{L}_{X}$ and $\mathcal D^{R}_{X}$ under the substitution \eqref{eq:prepoint}.
This anticipates their role on the forward/backward branches of the Keldysh contour in Fig.~\ref{fig:contour}: an operator acting on $\hat\rho$ from the left lives on the forward ($+$) branch and one acting from the right on the backward ($-$) branch, and the superscripts $L/R$ turn into the branch labels $\pm$.

Concatenating the $N$ slices, each carrying its own integration variables $(\Omega_n,\tilde\theta_n,\tilde\phi_n)$,
\begin{equation}
\begin{aligned}
\chi_N=\!\int\!\prod_{n=0}^{N-1}&\frac{d\Omega_n\,d\tilde\theta_n\,d\tilde\phi_n}{(2\pi)^2}\,
e^{\,i[\tilde\theta_n\Delta\theta_n+\tilde\phi_n\Delta\phi_n]}\\
&\times\bigl[1+\delta t\,\ell(\Omega_n;\tilde\theta_n,\tilde\phi_n)\bigr]\,\chi_0,
\end{aligned}
\end{equation}
and using $1+\delta t\,x=e^{\delta t\,x}+\mathcal O(\delta t^2)$, the weight of a path is gathered into a single exponential $e^{iS}$,
\begin{equation}
S=\int\!dt\,\bigl(\tilde\theta\,\dot\theta+\tilde\phi\,\dot\phi\bigr)
-i\!\int\!dt\;\ell(\Omega;\tilde\theta,\tilde\phi)=S_0+S_H+S_L.
\label{eq:Sab}
\end{equation}
Here $S_0$ is the first (kinetic) term, generated by the Fourier phases of the $\delta$-functions, while the generator \eqref{eq:ell-sided} splits into the Hamiltonian contribution, giving $S_H=-\!\int\!dt\,\bigl(H_+-H_-\bigr)$, and the sum over dissipation channels, giving $S_L$.
The coherent terms $S_0$ and $S_H$ are evaluated in Sec.~\ref{sec:coh} and the dissipative term $S_L$ in Sec.~\ref{sec:dissipative} for single-site decay.

\section{TWA description of a single spin}
\label{sec:singlespin}
The construction so far has produced an exact path integral for the Lindblad dynamics of a single spin-$1/2$, with the action \eqref{eq:Sab}. 
Using the Keldysh field theory formalism there is a well-developed route from such an action to the TWA: one rewrites the action in the so-called classical/quantum basis via a rotation from the $\pm$ basis and expands in the quantum fields.
Keeping terms up to quadratic order only in the quantum fields yields the semiclassical limit, which can be cast as a set of stochastic equations of motion \cite{kamenev2023field, sieberer2016keldysh}. 
In this section we first obtain such a semiclassical action for a single driven-decaying spin with detuning in Sec.~\ref{sec:semiclassical}, and then derive the TWA equations of motion from it in Sec.~\ref{sec:eom}.
The many-body interacting case will be discussed in Sec.~\ref{sec:manybody}.

\subsection{The semiclassical action}
\label{sec:semiclassical}
The action \eqref{eq:Sab} carries twice as many variables per slice as the sphere has coordinates, namely the configuration $(\theta, \phi)$ and the response pair $(\tilde \theta, \tilde \phi)$ introduced with the Fourier representation \eqref{eq:fourierdelta}. 
In Keldysh field theory, one also has a similar doubling of degrees of freedom, which has a physical origin: the density matrix is evolved from both sides, so every degree of freedom appears once on the forward and once on the backward branch of the closed time contour (Fig.~\ref{fig:contour}). 
We note that such a two-sided structure of the Lindbladian is still present in our construction, and it is stored in the sidedness of the operators in \eqref{eq:kernel-lindblad}, and more specifically in the $\pm$ labels of \eqref{eq:ell-sided}.
We now make the correspondence explicit, and also justify the identification of the response pair from \eqref{eq:fourierdelta}.

To this end, we examine how the forward and backward symbols $X_\pm$ depend on the response pair $(\tilde\theta,\tilde\phi)$.
At $\tilde\theta=\tilde\phi=0$, the forward and backward symbols coincide $X_\pm|_{\tilde\theta=\tilde\phi=0}=\operatorname{tr}[\hat X\hat\Delta(\Omega_n)]=X(\Omega_n)$, and so the difference $X_+-X_-$ is at least linear in $(\tilde\theta,\tilde\phi)$.
For spin-$1/2$ it is exactly linear, as we now show for the Hamiltonian term of \eqref{eq:ell-sided}, which by \eqref{eq:kernel-lindblad} represents the commutator,
$(\mathcal D^{L}_{H}-\mathcal D^{R}_{H})\hat\Delta=[\hat H,\hat\Delta]$.
Writing $\hat H=h_0\mathbb 1+\mathbf h\cdot\boldsymbol{\hat\sigma}$, the Pauli algebra gives
$[\hat H,\hat\Delta]
=\tfrac{\sqrt3}{2}\,[\mathbf h\cdot\boldsymbol{\hat\sigma},
\mathbf n\cdot\boldsymbol{\hat\sigma}]
=\sqrt3\,i\,(\mathbf h\times\mathbf n)\cdot\boldsymbol{\hat\sigma}$.
The cross product is orthogonal to $\mathbf n$ and therefore lies in the plane spanned by the tangent-frame vectors
$\mathbf e_\theta=\partial_\theta\mathbf n$ and
$\mathbf e_\phi=(\sin\theta)^{-1}\partial_\phi\mathbf n$,
with 
$\mathbf h\times\mathbf n
=(\mathbf h\cdot\mathbf e_\phi)\,\mathbf e_\theta
-(\mathbf h\cdot\mathbf e_\theta)\,\mathbf e_\phi$.
Moreover, every factor on the right is an angular derivative in disguise; differentiating the kernel and the symbol $H=h_0+\sqrt3\,\mathbf h\cdot\mathbf n$ gives
\begin{equation}
\begin{aligned}
\partial_\theta\hat\Delta&=\tfrac{\sqrt3}{2}\,\mathbf e_\theta\!\cdot\!\boldsymbol{\hat\sigma},
&\quad
\partial_\phi\hat\Delta&=\tfrac{\sqrt3}{2}\sin\theta\,\mathbf e_\phi\!\cdot\!\boldsymbol{\hat\sigma},\\
\partial_\theta H&=\sqrt3\,\mathbf h\cdot\mathbf e_\theta,
&\quad
\partial_\phi H&=\sqrt3\,\sin\theta\,\mathbf h\cdot\mathbf e_\phi .
\label{eq:commutator-id}
\end{aligned}
\end{equation}
Substituting these,
\begin{equation}
[\hat H,\hat\Delta]
=\frac{2i}{\sqrt3\,\sin\theta}
\bigl(\partial_\phi H\,\partial_\theta\hat\Delta
-\partial_\theta H\,\partial_\phi\hat\Delta\bigr).
\label{eq:spinbracket}
\end{equation}
Under the substitution \eqref{eq:prepoint} this becomes
\begin{equation}
H_+-H_-=-\frac{2}{\sqrt3\,\sin\theta}
\bigl(\tilde\phi\,\partial_\theta H-\tilde\theta\,\partial_\phi H\bigr).
\label{eq:sided-split}
\end{equation}

It is now convenient to introduce a change of variables.
We keep the propagated configuration, renaming it the classical field, and trade the response pair for the quantum fields,
\begin{equation}
\theta_{\rm cl}=\theta,\quad
\phi_{\rm cl}=\phi,\quad
\theta_q=-\frac{2\tilde\phi}{\sqrt3\,\sin\theta},\quad
\phi_q=\frac{2\tilde\theta}{\sqrt3\,\sin\theta},
\label{eq:clq-def}
\end{equation}
under which the difference becomes
\begin{equation}
H_+-H_-=\theta_q\,\partial_\theta H+\phi_q\,\partial_\phi H .
\label{eq:sided-split-q}
\end{equation}
With this, we can bring \eqref{eq:sided-split} into the canonical form it takes in Keldysh field theory.
The prefactor and the exchange of the components in \eqref{eq:clq-def} are read off from \eqref{eq:sided-split}: they are precisely what is needed for the difference to take the form \eqref{eq:sided-split-q} with unit coefficients.
Furthermore, the names are justified by the branch reading of this result. 
The standard Keldysh rotation trades the two copies on the forward and backward branches for their mean and their difference, 
\begin{equation}
\theta_\pm=\theta_{\rm cl}\pm\tfrac{\theta_q}{2},\qquad
\phi_\pm=\phi_{\rm cl}\pm\tfrac{\phi_q}{2},
\label{eq:pairing}
\end{equation}
the classical and the quantum field \cite{kamenev2023field}.
The classical field is the combination that can acquire an expectation value; it tracks the physical trajectory, and observables are evaluated on it. 
The quantum field measures the difference between the forward and the backward history and is forced to zero at the final end of the contour, where the trace sews the branches together.

\subsubsection{Coherent terms}
\label{sec:coh}
In fact, Taylor-expanding the symbol $H(\theta_\pm,\phi_\pm)$ about the classical configuration, $H(\theta_{\rm cl},\phi_{\rm cl})\equiv H_{\rm cl}$, using $\delta\equiv\theta_q\,\partial_\theta+\phi_q\,\partial_\phi$
for brevity, we get
\begin{equation}
H(\theta_\pm,\phi_\pm)
= H_{\rm cl}
\pm\tfrac12\,\delta H_{\rm cl}
+\tfrac18\,\delta^2 H_{\rm cl}
\pm\tfrac1{48}\,\delta^3 H_{\rm cl}
+\cdots .
\label{eq:H-taylor}
\end{equation}
In the difference of the two copies the even orders cancel while the
odd orders double,
\begin{equation}
H(\theta_+,\phi_+)-H(\theta_-,\phi_-)
=\delta H_{\rm cl}+\mathcal O(q^3),
\end{equation}
which is precisely the $\pm$ difference \eqref{eq:sided-split-q} to leading order in quantum fields with the Hamiltonian acting from the left evaluated on the forward configuration, and the one acting from the right on the backward one.
Therefore, the terms descending from the Hamiltonian contribution of the Lindbladian read
\begin{equation}
S_H^{\rm cl/q}
=-\!\int dt\,\bigl(\theta_q\,\partial_\theta H_{\rm cl}
+\phi_q\,\partial_\phi H_{\rm cl}\bigr).
\label{eq:SH-clq}
\end{equation}
Moreover, by inverting \eqref{eq:clq-def}, and substituting into \eqref{eq:Sab}, we obtain the kinetic contribution $S_0$, the first term in \eqref{eq:Sab},
\begin{equation}
S_0=\tfrac{\sqrt3}{2}\!\int\!dt\,\sin\theta_{\rm cl}\,
\bigl(\phi_q\dot\theta_{\rm cl}-\theta_q\dot\phi_{\rm cl}\bigr).
\label{eq:Stotal}
\end{equation}

The kinetic term $S_0$ and the single-spin Hamiltonian difference \eqref{eq:sided-split-q} are therefore both exactly linear in the quantum fields: there is no quadratic piece.
In the combined coherent action at linear order in the quantum fields, $S_0$ is the term carrying the time derivatives $\dot\theta_{\rm cl},\dot\phi_{\rm cl}$, which will form the left-hand side of the equations of motion, while $S_H$ contributes the drift $\partial_\theta H_{\rm cl},\partial_\phi H_{\rm cl}$ on the right-hand side. 
The coherent sector therefore contributes only to the drift and generates no noise; every noise term in the final stochastic equations must come from the dissipator.
To spell out the terms precisely for the driven spin with detuning,
\begin{equation}
H=\frac{\Omega_{\mathrm R}}{2}\,\hat\sigma^x+\frac{\Delta}{2}\hat\sigma^z,
\end{equation}
we first compute the symbols, which follow from the definition \eqref{eq:symbol} applied to each Pauli operator. 
Using the kernel \eqref{eq:swkernel} together with $\operatorname{tr}\hat\sigma^a=0$ and $\operatorname{tr}[\hat\sigma^a\hat\sigma^b]=2\delta^{ab}$, each Pauli matrix maps to the corresponding component of the Bloch vector,
\begin{equation}
\operatorname{tr}[\hat\Delta(\Omega)\,\hat\sigma^a]
=\tfrac12\operatorname{tr}\hat\sigma^a
+\tfrac{\sqrt3}{2}\sum_b n_b\operatorname{tr}[\hat\sigma^b\hat\sigma^a]
=\sqrt3\,n^a,
\label{eq:pauli-symbol}
\end{equation}
so that, with $\mathbf n$ from \eqref{eq:nvector},
\begin{equation}
\begin{aligned}
\hat\sigma^x&\mapsto\sqrt3\sin\theta\cos\phi,\\
\hat\sigma^y&\mapsto\sqrt3\sin\theta\sin\phi,\\
\hat\sigma^z&\mapsto\sqrt3\cos\theta.
\end{aligned}
\label{eq:pauli-symbols}
\end{equation}
Then, we have
\begin{equation}
H(\theta,\phi)=\sqrt3\bigl[\tfrac{\Omega_{\mathrm R}}{2}\sin\theta\cos\phi
+\tfrac{\Delta}{2}\cos\theta\bigr],
\end{equation}
whose coefficients in \eqref{eq:SH-clq} are
\begin{equation}
\partial_\theta H_{\rm cl}
=\sqrt3\,\tfrac{\Omega_{\mathrm R}}{2}\cos\theta_{\rm cl}\cos\phi_{\rm cl}
-\tfrac{\sqrt3\,\Delta}{2}\sin\theta_{\rm cl},
\label{eq:dH}
\end{equation}
and
\begin{equation}
\partial_\phi H_{\rm cl}
=-\sqrt3\,\tfrac{\Omega_{\mathrm R}}{2}\sin\theta_{\rm cl}\sin\phi_{\rm cl}.
\end{equation}

\subsubsection{Dissipative terms}
\label{sec:dissipative}

The dissipative part of the generator symbol \eqref{eq:ell-sided}, which enters the action through $S_L=-i\!\int\!dt\,\ell_D$, reads
\begin{equation}
\ell_D=\sum_\alpha\Bigl[\mathcal R_\alpha
-\tfrac12\bigl(L_\alpha^{\dagger}L_\alpha\bigr)_+
-\tfrac12\bigl(L_\alpha^{\dagger}L_\alpha\bigr)_-\Bigr],
\label{eq:ellD-def}
\end{equation}
mirroring the Lindblad dissipator \eqref{eq:lindblad} term by term, with left factors on the forward and right factors on the backward branch, as anticipated in Fig.~\ref{fig:contour}.
Here, $\mathcal R_\alpha$ denotes the polynomial of the composition $\mathcal D^{L}_{L_\alpha}\mathcal D^{R}_{L^\dagger_\alpha}$ under the substitution \eqref{eq:prepoint}.

We now evaluate $\ell_D$ for the decay channel $\hat L=\sqrt\Gamma\,\hat \sigma^-$ in the same classical/quantum basis.
Following \eqref{eq:pauli-symbols}, $\hat \sigma^\pm=\tfrac12(\hat \sigma^x\pm i\hat \sigma^y)$ has the symbol
\begin{equation}
\sigma^\pm\mapsto\tfrac{\sqrt3}{2}\sin\theta\,e^{\pm i\phi},
\end{equation}
so that 
\begin{equation}
L(\theta,\phi)=\tfrac{\sqrt{3\Gamma}}{2}\sin\theta\,e^{-i\phi},
\quad
L^*(\theta,\phi)=\tfrac{\sqrt{3\Gamma}}{2}\sin\theta\,e^{+i\phi},
\label{eq:Lsymbol}
\end{equation}
with $L_\pm\equiv L(\theta_\pm,\phi_\pm)$ and $L^*_\pm$ the symbol of $\hat L^\dagger=\sqrt\Gamma\,\hat \sigma^+$ on each branch. 
Unlike the coherent sector, the dissipator is not odd in the quantum fields: the no-jump terms enter as the symmetric sum forward plus backward and the jump term couples both branches, so $S_L$ carries a noise piece quadratic in the quantum fields, as well as the drift piece, linear in quantum fields.
It is also the place where the star product will make a difference to the underlying equations of motion.

The exact evaluation keeps the identities \eqref{eq:onesided} in full, i.e.~the complete differential action of the jump operators on the kernel $\hat\Delta$. 
Explicitly, they read (see App.~\ref{app:diff-id})
\begin{equation}
\begin{aligned}
\hat\sigma^-\hat\Delta &=\tfrac12 e^{-i\phi}\Bigl[\sqrt3\sin\theta
+(\sqrt3\cos\theta+1)\partial_\theta\\
&\hphantom{{}=\tfrac12 e^{-i\phi}\Bigl[}
-i\bigl(\tfrac{\csc\theta}{\sqrt3}+\cot\theta\bigr)\partial_\phi
+\tfrac{2\csc\theta}{\sqrt3}\partial_\phi^2\Bigr]\hat\Delta,\\[4pt]
\hat\Delta\,\hat\sigma^+ &=\tfrac12 e^{+i\phi}\Bigl[\sqrt3\sin\theta
+(\sqrt3\cos\theta+1)\partial_\theta\\
&\hphantom{{}=\tfrac12 e^{+i\phi}\Bigl[}
+i\bigl(\tfrac{\csc\theta}{\sqrt3}+\cot\theta\bigr)\partial_\phi
+\tfrac{2\csc\theta}{\sqrt3}\partial_\phi^2\Bigr]\hat\Delta,
\end{aligned}
\label{eq:diffident}
\end{equation}
with the two phases $e^{\mp i\phi}$ carried by the left-acting $\hat\sigma^-$ and the right-acting $\hat\sigma^+$ being what survives as the relative phase of the two branches.
For a two-level system the jump term can be written as
$\hat\sigma^-\hat X\hat\sigma^+ =\operatorname{tr}\bigl[\hat\sigma^+\hat\sigma^-\hat X\bigr]\, \hat\sigma^-\hat\sigma^+$ for any $\hat X$, so that $\hat\sigma^-\hat\Delta\,\hat\sigma^+ =\tfrac12\bigl(1+\sqrt3\cos\theta\bigr)\,\hat\sigma^-\hat\sigma^+$ is a single $2\times2$ matrix. Expanding it in the basis \eqref{eq:der-basis} of the kernel (App.~\ref{app:recycling}), we get 
\begin{equation}
\begin{aligned}
\hat\sigma^-\hat\Delta\,\hat\sigma^+
={}&\tfrac12\bigl(1+\sqrt3\cos\theta\bigr)\hat\Delta
+\frac{(1+\sqrt3\cos\theta)^2}{2\sqrt3\,\sin\theta}\,\partial_\theta\hat\Delta\\
&+\frac{(1+\sqrt3\cos\theta)(\sqrt3+\cos\theta)}{2\sqrt3\,\sin^2\theta}\,
\partial_\phi^2\hat\Delta.
\end{aligned}
\label{eq:rec-expansion}
\end{equation}
For the no-jump matrix $\tfrac12\{\hat\sigma^+\hat\sigma^-,\hat\Delta\}$, we get
\begin{equation}
\begin{aligned}
\tfrac12\{\hat\sigma^+\hat\sigma^-,\hat\Delta\}
={}&\tfrac12\bigl(1+\sqrt3\cos\theta\bigr)\hat\Delta
-\frac{1-3\cos^2\theta}{2\sqrt3\,\sin\theta}\,\partial_\theta\hat\Delta\\
&+\frac{\cos\theta}{\sqrt3\,\sin^2\theta}\,\partial_\phi^2\hat\Delta.
\end{aligned}
\label{eq:nj-expansion}
\end{equation}
The two $\hat\Delta$ coefficients are equal, as they should be: the coefficient of $\hat\Delta$ in the basis expansion of any matrix is its trace, since $\operatorname{tr}\hat\Delta=1$, while the derivatives of the kernel are traceless, and the two traces coincide by cyclicity, $\operatorname{tr}[\hat\sigma^-\hat\Delta\hat\sigma^+] =\operatorname{tr}[\hat\sigma^+\hat\sigma^-\hat\Delta]$.
Since the jump and no-jump entries carry opposite signs in the dissipator, the $\hat\Delta$ terms cancel identically, and the dissipator applied to the kernel consists of derivative terms only, which corresponds to pure drift and diffusion.
Altogether,
\begin{equation}
\Gamma\Bigl[\hat\sigma^-\hat\Delta\,\hat\sigma^+
-\tfrac12\{\hat\sigma^+\hat\sigma^-,\hat\Delta\}\Bigr]
=A_\theta\,\partial_\theta\hat\Delta
+\tfrac12 B_{\phi\phi}\,\partial_\phi^2\hat\Delta,
\label{eq:kernel-diss}
\end{equation}
with
\begin{equation}
A_\theta=\Gamma\Bigl(\cot\theta+\tfrac{\csc\theta}{\sqrt3}\Bigr)
\label{eq:fp-coeffs1}
\end{equation}
and 
\begin{equation}
B_{\phi\phi}=\Gamma\Bigl(1+2\cot^2\theta
+\tfrac{2\cot\theta\csc\theta}{\sqrt3}\Bigr).
\label{eq:fp-coeffs2}
\end{equation}
The substitution \eqref{eq:prepoint} then yields the dissipative symbol,
\begin{equation}
\ell_D=-A_\theta\,i\tilde\theta+\tfrac12 B_{\phi\phi}\,(i\tilde\phi)^2 .
\label{eq:ellD-star}
\end{equation}
Equation \eqref{eq:ellD-star} is in fact exact, as the basis \eqref{eq:der-basis} of a single spin-$1/2$ carries at most two derivatives, so the expansion \eqref{eq:kernel-diss} terminates for any channel, and the dissipative sector is at most quadratic in the response/quantum fields.
Indeed, read in continuous time, the update \eqref{eq:stepweight} with \eqref{eq:ellD-star} is a Fokker--Planck equation\footnote{Because the $\hat\Delta(\Omega)$ are linearly dependent, the weight $\chi$ representing a given state is not unique, as elaborated on the gauge freedom of $\mathrm{SU}(2)$ phase-space representations in Refs.~\cite{mink2022hybrid,hartmann2026truncatedwignerapproximationspins}.
Equation \eqref{eq:fp} propagates one valid choice of weight, which for $t>0$ is no longer $\sin\theta\,W/2\pi$ itself; all observables obtained
from it are nevertheless exact, cf.\ the moment closure below \eqref{eq:eom}.} of the form
\begin{equation}
\partial_t\chi=-\partial_\theta\bigl(A_\theta\,\chi\bigr)
+\tfrac12\,\partial_\phi^2\bigl(B_{\phi\phi}\,\chi\bigr).
\label{eq:fp}
\end{equation}
The drift vanishes at $\cos\theta^*=-1/\sqrt3$, the latitude of the radius-$\sqrt3$ sphere on which the symbol of $\hat\sigma^z$ equals $\sqrt3\cos\theta^*=-1$.
Equation \eqref{eq:fp} coincides, after the convention map $\theta\to\pi-\theta$, $\phi\to-\phi$, with the continuous TWA result of Ref.~\cite{mink2022hybrid}.
Inverting \eqref{eq:clq-def} and computing $S_L=-i\!\int\!dt\,\ell_D$, the  $i\tilde\theta$ entry becomes the real (drift) part of the action, and the $(i\tilde\phi)^2$ entry the imaginary (noise) part,
\begin{equation}
\begin{aligned}
S_L^{(\star)}
={}&-\!\int\! dt\,\phi_q\,\frac{\Gamma}{2}\bigl(1+\sqrt3\cos\theta_{\rm cl}\bigr)\\
&+\frac{3i\Gamma}{8}\!\int\! dt\,\theta_q^{2}
\Bigl[1+\cos^2\theta_{\rm cl}+\tfrac{2}{\sqrt3}\cos\theta_{\rm cl}\Bigr].
\end{aligned}
\label{eq:SL-star}
\end{equation}
If instead we had evaluated the dissipative terms with the ordinary product of the branch symbols in place of the star product (the step that within the present representation mirrors the truncation route of the bosonic constructions \cite{sieberer2016keldysh, hosseinabadi2025user}), the resulting action, derived in App.~\ref{sec:app-multi}, would display incorrect terms in both sectors: an incorrect drift $-\tfrac{3\Gamma}{4}\sin^2\theta_{\rm cl}$ that drives every trajectory towards the south pole of the sphere, and a spurious $\phi_q^2$ noise channel emerges.
For certain types of dissipation, however, the additional terms generated by the star product vanish identically, and the ordinary product happens to give the correct action. 
This is the case for dephasing, modelled by the Hermitian jump operator $\hat L=\sqrt{\gamma}\,\hat\sigma^z$: the dissipator reduces to a double commutator, and both evaluations map it onto the same purely diffusive noise term, as we show in App.~\ref{app:multi-deph}.
This is also why the TWA simulations of the dissipative Rydberg gas in Ref.~\cite{noel2026quantum} could proceed without the star product: the dissipation considered there is dephasing, so the ordinary product of symbols already generates the correct dissipative action, and no corrections are missed.

\subsection{Equations of motion}
\label{sec:eom}
The full action assumes the canonical form of an MSRJD stochastic field theory: terms linear in the quantum fields, whose coefficients will become the deterministic equations of motion, plus a quadratic form in the quantum fields, whose coefficient matrix is the noise kernel. 
For the driven-decaying spin only the $\theta_q^2$ entry of this matrix survives, giving
\begin{equation}
S=\int\!dt\,\Bigl[\theta_q\,\mathcal E_\theta
+\phi_q\,\mathcal E_\phi
+\tfrac{i}{2}\,D\,\theta_q^{2}\Bigr],
\label{eq:S-canonical}
\end{equation}
with
\begin{equation}
\begin{aligned}
\mathcal E_\theta&=-\tfrac{\sqrt3}{2}\sin\theta_{\rm cl}\,\dot\phi_{\rm cl}
-\partial_\theta H_{\rm cl},\\
\mathcal E_\phi&=\tfrac{\sqrt3}{2}\sin\theta_{\rm cl}\,\dot\theta_{\rm cl}
-\partial_\phi H_{\rm cl}
-\tfrac{\Gamma}{2}\bigl(1+\sqrt3\cos\theta_{\rm cl}\bigr),\\
D&=\tfrac{3\Gamma}{4}\Bigl[1+\cos^2\theta_{\rm cl}
+\tfrac{2}{\sqrt3}\cos\theta_{\rm cl}\Bigr]
=\tfrac34\sin^2\theta_{\rm cl}\,B_{\phi\phi}.
\end{aligned}
\label{eq:EED}
\end{equation}
Note the crosswise pairing in \eqref{eq:EED}: the coefficient of $\theta_q$ contains $\dot\phi_{\rm cl}$ and $\partial_\theta H_{\rm cl}$, while that of $\phi_q$ contains $\dot\theta_{\rm cl}$ and $\partial_\phi H_{\rm cl}$, as expected for canonically conjugate angles.
In particular, the noise, which couples to $\theta_q$, will enter the equation of motion for $\phi_{\rm cl}$.
Since the coherent sector is exactly linear and the dissipative sector \eqref{eq:SL-star} exactly quadratic in the quantum fields, there is nothing to truncate: \eqref{eq:S-canonical} is the full action of the single driven-decaying spin, and every step that follows is exact.
The weight $e^{iS}$ in the path integral now reads
\begin{equation}
e^{iS}
=\exp\Bigl(i\!\int dt\bigl[\theta_q\mathcal E_\theta
+\phi_q\mathcal E_\phi\bigr]\Bigr)
\exp\Bigl(i\frac{i}{2}\!\int\!dtD\theta_q^{2}\Bigr).
\label{eq:weight-split}
\end{equation}
The quadratic term can be represented by a Gaussian identity.
Since in the noise kernel, only the $\theta_q^2$ entry is
nonzero, a single auxiliary real Gaussian (Hubbard--Stratonovich) field $\xi(t)$  suffices,
\begin{equation}
\exp\Bigl(-\frac12\!\int\!dt\,D\,\theta_q^{2}\Bigr)
\propto\int\!\mathcal D[\xi]\,
\exp\Bigl(-\frac12\!\int\!dt\,\frac{\xi^{2}}{D}
-i\!\int\!dt\,\xi\,\theta_q\Bigr),
\label{eq:HS}
\end{equation}
and the action becomes linear in the quantum fields
\begin{equation}
S=\int\!dt\,\bigl[\theta_q\,\bigl(\mathcal E_\theta-\xi\bigr)
+\phi_q\,\mathcal E_\phi\bigr]\;+\;S_\xi
\label{eq:S-linear}
\end{equation}
with
\begin{equation}
S_\xi=\frac{i}{2}\int\!dt\,\frac{\xi^{2}}{D},
\label{eq:Sxi}
\end{equation}
which implies
\begin{equation}
\langle\xi(t)\,\xi(t')\rangle=D(\theta_{\rm cl})\,\delta(t-t').
\label{eq:noise-corr}
\end{equation}
The action is then
\begin{equation}
S=\int dt \Bigl\{\tfrac{\sqrt3}{2}\sin\theta_{\rm cl}
\Bigl[\phi_q\bigl(\dot\theta_{\rm cl}-F_\theta\bigr)
-\theta_q\bigl(\dot\phi_{\rm cl}-F_\phi\bigr)\Bigr]
-\theta_q\,\xi\Bigr\}+S_\xi,
\label{eq:S-msr-clq}
\end{equation}
with 
\begin{equation}
\begin{aligned}
F_\theta&=\frac{2}{\sqrt3\sin\theta_{\rm cl}}
\Bigl[\partial_\phi H_{\rm cl}
+\tfrac{\Gamma}{2}\bigl(1+\sqrt3\cos\theta_{\rm cl}\bigr)\Bigr]\\
&=-\Omega_{\mathrm R}\sin\phi_{\rm cl}
+\Gamma\Bigl(\cot\theta_{\rm cl}
+\tfrac{\csc\theta_{\rm cl}}{\sqrt3}\Bigr),\\[2pt]
F_\phi&=-\frac{2}{\sqrt3\sin\theta_{\rm cl}}\,
\partial_\theta H_{\rm cl}
=\Delta-\Omega_{\mathrm R}\cot\theta_{\rm cl}\cos\phi_{\rm cl}.
\end{aligned}
\label{eq:drifts}
\end{equation}
The functional integral over the quantum fields can now be performed explicitly,
\begin{equation}
\begin{aligned}
\int\!\mathcal D[\theta_q]\,\mathcal D[\phi_q]
&\exp\Bigl(i\!\int\!dt\,\bigl[\theta_q(\mathcal E_\theta-\xi)
+\phi_q\,\mathcal E_\phi\bigr]\Bigr)\\
&\propto\;\delta\bigl[\mathcal E_\theta-\xi\bigr]\,
\delta\bigl[\mathcal E_\phi\bigr].
\end{aligned}
\label{eq:functional-delta}
\end{equation}
The $\delta$-functionals restrict the classical configurations to the solutions of $\mathcal E_\phi=0$ and $\mathcal E_\theta=\xi$, and the partition function reduces to an average over trajectories obeying these two equations. 

The remaining ingredient is the sampling of the initial state.
Since the continuous symbol of a pure state is not positive, it cannot be sampled directly as a probability distribution for the initial conditions \cite{mink2022hybrid}.
However, the $\mathrm{SU}(2)$ Wigner function is not unique, and a gauge transformation can render the initial distribution positive \cite{mink2022hybrid,hartmann2026truncatedwignerapproximationspins}.
For eigenstates of the Pauli operators, such a positive representative is given by the discrete Wigner representation \cite{wootters1987wigner}, from which we sample the initial state.
For states polarised along a coordinate axis, it reproduces all first and symmetrically ordered second moments exactly.
Solving the constraint equations for the time derivatives with \eqref{eq:drifts} gives
\begin{equation}
\dot\theta_{\rm cl}=F_\theta,\qquad
\dot\phi_{\rm cl}=F_\phi-\frac{2}{\sqrt3\sin\theta_{\rm cl}}\,\xi(t),
\label{eq:eom-raw}
\end{equation}
driven by the Gaussian noise \eqref{eq:noise-corr}.
By \eqref{eq:EED} and \eqref{eq:noise-corr} the rescaled noise has correlator 
\begin{equation}
\bigl\langle\tfrac{2\,\xi(t)}{\sqrt3\sin\theta_{\rm cl}}\,
\tfrac{2\,\xi(t')}{\sqrt3\sin\theta_{\rm cl}}\bigr\rangle
=\tfrac{4}{3\sin^2\theta_{\rm cl}}\,D\,\delta(t-t')
=B_{\phi\phi}\,\delta(t-t')
\end{equation}
so in terms of unit-strength white noise the stochastic equations of
motion of the driven-decaying spin follow as
\begin{equation}
\begin{aligned}
\dot\theta_{\rm cl}&=-\Omega_{\mathrm R}\sin\phi_{\rm cl}
+\Gamma\Bigl(\cot\theta_{\rm cl}+\tfrac{\csc\theta_{\rm cl}}{\sqrt3}\Bigr),\\
\dot\phi_{\rm cl}&=\Delta-\Omega_{\mathrm R}\cot\theta_{\rm cl}\cos\phi_{\rm cl}
+\sqrt{B_{\phi\phi}}\;\xi(t),
\end{aligned}
\label{eq:eom}
\end{equation}
with $\langle\xi(t)\xi(t')\rangle=\delta(t-t')$, and $B_{\phi\phi}$ from
\eqref{eq:fp-coeffs2}.

Two remarks complete the single-spin analysis.
First, a stochastic equation with multiplicative noise is not fully specified until a discretisation convention is chosen; here, since the noise amplitude depends only on $\theta_{\rm cl}$ while the noise acts on $\phi_{\rm cl}$, the It\^o and Stratonovich readings of \eqref{eq:eom} coincide, and the equations may be integrated without further specification.
Second, the equations pass an exact benchmark: the It\^o moment closure of \eqref{eq:eom} on the observables $\sqrt3\,\mathbf n$ reproduces the exact equations of motion for $\langle\hat\sigma^a\rangle$ that follow from the master equation \eqref{eq:lindblad}, including all $\Omega_{\mathrm R},\Delta,\Gamma$ cross terms, in agreement with Ref.~\cite{mink2022hybrid}.
The star-product TWA thus captures the noninteracting spin exactly at the level of first moments, and approximation errors can only enter through the interaction terms, to which we now turn.

\section{TWA with interacting spins}
\label{sec:manybody}

So far, the construction was carried out for a single spin.
We now repeat it for $N$ interacting spins and record what changes. 
As a concrete example we take the driven-dissipative Rydberg array, where each spin is subject to the drive $\Omega_{\rm R}$ and the spins are coupled by a density-density interaction \cite{Weimer_2010, Saffman_2010, Browaeys_2020},
\begin{equation}
H=\sum_i\left(\frac{\Omega_{\rm R}}{2} \hat{\sigma}_i^x+\frac{\Delta}{2} \hat{\sigma}_i^z\right)+\sum_{i<j} V_{i j} \hat{n}_i \hat{n}_j
\label{eq:ryd-ham}
\end{equation}
with $\hat n _i = \hat \sigma _i^{+} \hat \sigma _i ^{-}$.
For the dissipation we consider local decay, $\hat L_i=\sqrt{\Gamma}\,\hat\sigma_i^{-}$, acting independently on each site.
We start with the many-body version of the kernel \eqref{eq:swkernel}, which is the tensor product of single-site kernels, 
\begin{equation}
\hat\Delta(\Omega_1, \Omega_2, \ldots,\Omega_N)=\bigotimes_{i=1}^N \hat\Delta\left(\Omega_i\right),
\label{eq:manybody-sw}
\end{equation}
and the appropriately generalised symbol map \eqref{eq:symbol} carries over with the measure $d\mu \rightarrow \prod_i d\mu_i$, normalised to $2^N$.
As a consequence, for operators on different sites, the symbol of the product is the product of the symbols $\operatorname{tr}\left[\hat\Delta(\Omega) \hat{X}_i \hat{Y}_j\right]=X\left(\Omega_i\right) Y\left(\Omega_j\right)$ for $i \neq j$, so the star product is nontrivial only for products of symbols referring to the same site. 
Then, the many-body theory differs only in the interaction terms of the Hamiltonian. 
The construction of the path integral in Sec.~\ref{sec:construction} goes through verbatim, generalised to the many-body case.
The dissipative sector does not change at all either, with each jump operator acting on a single factor of \eqref{eq:manybody-sw}, the identities \eqref{eq:Lsymbol}-\eqref{eq:nj-expansion} apply in the variables of site $i$, and the dissipative action will be a sum of exact per-site copies.\footnote{For collective jump operators, which would couple sites in the dissipator, see Ref.~\cite{Mink_2023}.}

Qualitatively new terms arise from the interaction term, as it is the only term that couples two sites.
Concretely, the interaction acts on two factors of the many-body kernel \eqref{eq:manybody-sw} at once, each carrying its own pair of angles $\Omega_i=(\theta_i,\phi_i)$.
We evaluate these by first splitting the action of $\hat n$ on its site's kernel into its symmetric $\mathcal S$ and antisymmetric $\mathcal C$ parts by considering 
\begin{equation}
\hat{n} \hat\Delta=(\mathcal{S}+i \mathcal{C}) \hat\Delta, \quad \hat\Delta \hat{n}=(\mathcal{S}-i \mathcal{C}) \hat\Delta.
\label{eq:split}
\end{equation}
The split is merely the sum and difference of the two identities \eqref{eq:onesided}.
Adding and subtracting \eqref{eq:split} gives $\mathcal S \hat\Delta = \tfrac{1}{2} \{ \hat n, \hat\Delta\}$ and $i \mathcal C \hat\Delta = \tfrac{1}{2}  [\hat n, \hat\Delta]$, so that the symmetric part is the anticommutator half and the antisymmetric part the commutator half by construction. 
Then it can be easily seen that the symmetric part of \eqref{eq:split} is simply the no-jump expansion we have already computed in \eqref{eq:nj-expansion}, since $\hat n = \hat \sigma ^{+} \hat \sigma^{-} \propto \hat L^{\dagger} \hat L$.
Therefore, we can read off the coefficients from \eqref{eq:nj-expansion} to get  $\mathcal{S}=n(\theta)+b(\theta) \partial_\theta+d(\theta) \partial_\phi^2$
with 
\begin{equation}
b(\theta)=-\frac{1-3 \cos ^2 \theta}{2 \sqrt{3} \sin \theta}, \quad d(\theta)=\frac{\cos \theta}{\sqrt{3} \sin ^2 \theta},
\end{equation}
and with $n(\theta) = \tfrac{1}{2} (1+\sqrt3  \cos(\theta))$ being the symbol of $\hat n$. 
The antisymmetric part follows from the commutator identity \eqref{eq:commutator-id} applied to this symbol $\tfrac{1}{2} [\hat n, \hat\Delta]=\tfrac{i}{2}\partial_{\phi} \hat\Delta$, so that $\mathcal C = \tfrac{1}{2} \partial_\phi$. 

To evaluate the commutator, we apply \eqref{eq:split} once per site: we start with $\hat n_j\hat\Delta=(\mathcal S_j+i\mathcal C_j)\hat\Delta$, which commutes with $\hat n_i$, since $\mathcal S_j,\mathcal C_j$ are differential operators in the angles of site $j$ only.
This can therefore be converted in turn, giving $\hat n_i\hat n_j\hat\Delta=(\mathcal S_i+i\mathcal C_i)(\mathcal S_j+i\mathcal C_j)\hat\Delta$ and analogously for $\hat\Delta\hat n_i\hat n_j$.
Subtracting the two orderings, the $\mathcal S\mathcal S$ and $\mathcal C\mathcal C$ terms cancel and we get
\begin{equation}
\left[\hat{n}_i \hat{n}_j, \hat\Delta\right]=2 i\left(\mathcal{C}_i \mathcal{S}_j+\mathcal{S}_i \mathcal{C}_j\right) \hat\Delta.
\end{equation}
Inserting the expressions for $\mathcal S, \mathcal C$ yields
\begin{equation}
\begin{aligned}
& {\left[\hat{n}_i \hat{n}_j, \hat\Delta\right]=i\left[n\left(\theta_j\right) \partial_{\phi_i}+b\left(\theta_j\right) \partial_{\phi_i} \partial_{\theta_j}\right.} \\
&\left.+d\left(\theta_j\right) \partial_{\phi_i} \partial_{\phi_j}^2+(i \leftrightarrow j)\right] \hat\Delta.
\label{eq:diff-id-int}
\end{aligned}
\end{equation}
The first entry is the expected drift, as each spin precesses in the field set by the symbols of its neighbours. 
However, the mixed-spin derivatives are new kinds of terms, which we did not encounter for the single-spin case, and would not encounter for bosons either.\footnote{For bosons, the symmetric part of the Moyal product carries only even and the antisymmetric part only odd derivative orders, and so a Hamiltonian generates only odd powers of the quantum fields, see \cite{polkovnikov2010phase}. The spin-$1/2$ anticommutator instead contains first derivatives, and then its product with the neighbour's commutator produces the mixed second-order term.} 
Then, under the Fourier resolution \eqref{eq:fourierdelta} and the rotation \eqref{eq:clq-def}, the three entries of \eqref{eq:diff-id-int} enter the action as 
\begin{equation}
\begin{aligned}
S_{\rm int}^{\rm cl/q}&=-\sum_{i<j}V_{ij}\!\int\! dt\,\Bigl\{
-\frac{\sqrt{3}}{2}\left[\sin \theta_{{\rm cl},i}\, n\left(\theta_{{\rm cl},j}\right)\, \theta_{q, i}+(i \leftrightarrow j)\right] \\
&-\frac{i \sqrt{3}}{8}\left[\sin \theta_{{\rm cl},i}\left(1-3 \cos ^2 \theta_{{\rm cl},j}\right) \theta_{q, i}\, \phi_{q, j}+(i \leftrightarrow j)\right] \\
&+\frac{3}{8}\left[\sin \theta_{{\rm cl},i} \cos \theta_{{\rm cl},j}\, \theta_{q, i}\, \theta_{q, j}^2+(i \leftrightarrow j)\right]\Bigr\}.
\label{eq:nn-orders}
\end{aligned}
\end{equation}

Previously, for the single spin in Sec.~\ref{sec:singlespin}, obtaining the stochastic trajectories from the action \eqref{eq:S-canonical} relied on the term quadratic in the quantum fields being decoupled by the Gaussian identity \eqref{eq:HS} to obtain the noise.
The remaining action, once linear in the quantum fields, gave the $\delta$-functional \eqref{eq:functional-delta} conditions, reducing the path integral to an average over trajectories.
The interaction spoils the structure that made these steps possible: the third line of \eqref{eq:nn-orders} is cubic in the quantum fields, and the quadratic terms in its second line cannot be traded for any real noise.
No term of the action \eqref{eq:S-canonical} or \eqref{eq:nn-orders} supplies a diagonal $\phi_q^2$ entry, so a pair of noises capable of generating these quadratic terms would have to satisfy $\langle\xi_\phi\xi_\phi\rangle=0$ together with $\langle\xi_\theta\xi_\phi\rangle=c\neq0$, where $c\propto\sin\theta_{{\rm cl},i}\,(1-3\cos^2\theta_{{\rm cl},j})$ is the coefficient of the mixed quadratic term. 
No \emph{real} random variable can be simultaneously of zero variance and correlated with another. 
A real trajectory ensemble therefore cannot accommodate the mixed second-derivative term, and as a result we discard all mixed-spin derivatives coming from the interaction term.
This step is not controlled by order counting: the second line of \eqref{eq:nn-orders} is quadratic in the quantum fields, the same order as the dissipative noise retained in \eqref{eq:SL-star}. 
It is also not controlled by a small parameter, as for spin-$1/2$ there is no $1/S$ to expand in.

Once the mixed-spin derivative terms are discarded, the remaining steps to obtain the equations of motion then mirror those of the single-spin derivation from Sec.~\ref{sec:eom}: one Hubbard--Stratonovich field is introduced per site as in Eq.~\eqref{eq:HS}, and integrating out the quantum fields enforces the constraint of the $\delta$-functional, Eq.~\eqref{eq:functional-delta}, site by site.
This leads to the coupled stochastic differential equations
\begin{equation}
\begin{aligned}
\dot{\theta}_{{\rm cl},i} &= -\Omega_{\mathrm R}\sin\phi_{{\rm cl},i}
+\Gamma\left(\cot\theta_{{\rm cl},i}+\frac{\csc\theta_{{\rm cl},i}}{\sqrt{3}}\right), \\
\dot{\phi}_{{\rm cl},i} &= \Delta+\sum_{j\neq i} V_{ij}\, n(\theta_{{\rm cl},j}) \\
  &\quad -\Omega_{\mathrm R}\cot\theta_{{\rm cl},i}\cos\phi_{{\rm cl},i}
  +\sqrt{B_{\phi\phi}(\theta_{{\rm cl},i})}\,\xi_i(t),
\end{aligned}
\label{eq:eom-many}
\end{equation}
with independent unit white noises $\left\langle\xi_i(t) \xi_j\left(t^{\prime}\right)\right\rangle=\delta_{i j} \delta\left(t-t^{\prime}\right)$, and $B_{\phi \phi}$ from \eqref{eq:fp-coeffs2}. 
By the structure of \eqref{eq:drifts}, the drift of $\theta_{{\rm cl},i}$ is generated by the $\phi_i$-derivative of the Hamiltonian and the drift of $\phi_{{\rm cl},i}$ by the $\theta_i$-derivative; since the interaction depends on the $\theta_{{\rm cl},i}$ only, it enters solely the equation for $\phi_{{\rm cl},i}$.

While the argument above explains why the mixed-derivative terms cannot be incorporated into the stochastic equations of motion, it does not bound the error made by discarding them: as noted above, the second line of \eqref{eq:nn-orders} is of the same order in the quantum fields as the dissipative noise we keep.
Discarding the third line, which is cubic in the quantum fields, is the standard TWA truncation of terms beyond quadratic order, but here no \textit{a priori} small parameter controls it either.
We therefore now identify the regimes in which the discarded terms are small compared to the interaction drift that is retained (first line of \eqref{eq:nn-orders}).
One regime in which the mixed-derivative terms may safely be discarded is that of a high effective coordination number, i.e.~an interaction that couples many spins with comparable strength, the cleanest case being the all-to-all limit $V_{ij}=V$.
For density-density interactions, the retained term enters the equations of motion \eqref{eq:eom-many} as the deterministic shift of the precession, $\sum_{j} V_{ij}\, n(\theta_{{\rm cl}, j})$.
Here, spin $j$ enters the equation of motion of spin $i$ only through the scalar coefficient $n(\theta_{{\rm cl}, j})$, while the derivative $\partial_{\phi_i}$ is the same operator for every $j$.
The sum over the $N-1$ coupled spins therefore adds up the coefficients into a single drift term of order $NV$.
In the discarded entries, by contrast, spin $j$ contributes a derivative, as in $\partial_{\phi_i}\partial_{\theta_j}$.
Derivatives with respect to different sites are distinct operators, so the sum over $j$ cannot pile up into one large term and there remain $N-1$ separate contributions, each of order $V$.
An explicit counting discussed in Ref.~\cite{mink2022hybrid}, appropriate for collective observables, shows the mixed second derivatives to be suppressed by $1/N$ and the mixed third derivatives by $1/N^{2}$ relative to the drift.
One can rationalise that the TWA improves with growing coordination number, since the spins coupled to a given site can be grouped into a local ensemble that effectively acts on it as a single macroscopic spin, whose fluctuations are small.
When the coordination number is instead of order one, as for the nearest-neighbour chain, there is no ensemble to group, and the discarded terms are no longer parametrically small.

Nevertheless, two further mechanisms mitigate the error even then.
First, all discarded terms are proportional to the interaction strength $V$, so for weak interactions they remain small compared to the single-spin terms, which the truncation treats exactly, and deviations can build up only over times of order $1/V$.
Second, the mixed second derivative is inactive at early times for initial states polarised along the $z$ axis, such as the ground state natural to a Rydberg array or the fully excited state used in Sec.~\ref{sec:results}.
Any sampling that reproduces the second moment $\langle(\sigma^z)^2\rangle=1$ alongside $\langle\sigma^z\rangle=\pm1$ places every trajectory on the latitude $\cos\theta_{\rm cl}=\pm1/\sqrt{3}$, on which the coefficient $b(\theta_j)\propto 1-3\cos^2\theta_j$ of the mixed term in \eqref{eq:diff-id-int} vanishes identically.

\section{Numerical simulations} 
\label{sec:results}

The preceding analysis leaves open how accurate the truncation is for the driven-dissipative Rydberg chain \eqref{eq:ryd-ham}. 
In order to obtain an understanding of the applicability of the TWA, we therefore benchmark the stochastic equations of motion derived above against exact solutions of the corresponding master equations.
First, for a single driven spin in the presence of decay, for which the equations are exact and only the phase-space mapping of the dissipation is tested; and then for the driven-dissipative Rydberg chain \eqref{eq:ryd-ham}.
The Rydberg chain is in fact a challenging benchmark: its coordination number is of order one, precisely the regime identified in Sec.~\ref{sec:manybody} in which the discarded interaction terms are not parametrically small.
Table~\ref{tab:cartesian-eom} summarises the TWA equations contribution by contribution in Cartesian components. 
The interaction row is the only approximate entry, every single-site row being treated exactly.
The central element of the construction is that the dissipative rows of the table follow from the $\mathrm{SU}(2)$ star product. 
Meanwhile, a naive transcription of the bosonic TWA to spins would instead multiply the symbols of the jump operators as ordinary functions, and the benchmarks below are designed to highlight the difference resulting from this replacement.
To isolate the importance of faithfully mapping operator products to the $\mathrm{SU}(2)$ phase-space in the dissipative sector, we compare two sets of trajectories. 
The first set follows the equations of Table~\ref{tab:cartesian-eom}, which we refer to as ``TWA with star product''.
For the second set the $\mathrm{SU}(2)$ star product is replaced by the ordinary multiplicative product (see App.~\ref{sec:app-multi} and Ref.~\cite{hosseinabadi2025user}), referred to as ``TWA w/o star product'', which corresponds to the naive transcription of the bosonic construction to spins.

\begin{figure}
    \centering
    \includegraphics[width=1.02\linewidth]{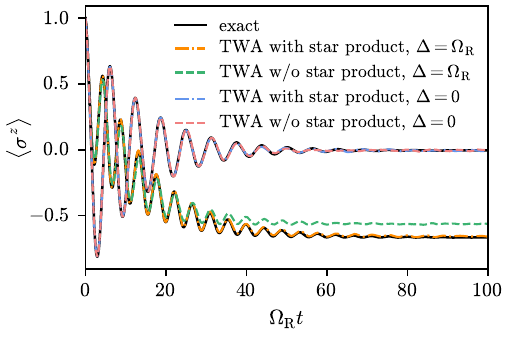}
\caption{\textbf{Single-spin benchmark.}
Site magnetisation $\langle\sigma^z\rangle$ of a single driven-decaying spin: exact solution of the master equation (black), the stochastic equations of Table~\ref{tab:cartesian-eom} (``TWA with star product''), and the variant with operator products replaced by ordinary multiplication of their symbols (``TWA w/o star product'') at detuning $\Delta = \Omega_{\mathrm R}$ and on resonance, $\Delta = 0$.
On resonance all curves coincide, while at finite detuning the ordinary-product variant relaxes to a visibly displaced steady state.}
    \label{fig:single-spin}
\end{figure}

We first consider a single driven-decaying spin.
The spin is initialised in the state $\ket{\uparrow}$ and evolved with Rabi frequency $\Omega_{\rm R}$ and decay with $\Gamma=0.1\,\Omega_{\mathrm R}$, both for resonant drive ($\Delta=0$) and at finite detuning ($\Delta=\Omega_{\rm R}$).
While the ground state $\ket{\downarrow}$ would be the natural starting point for a Rydberg array, it is left invariant by the decay, so no dissipative dynamics would occur until the drive has built up population. 
The excited state instead undergoes decay from the very beginning.
This initial state is sampled from its discrete Wigner representation: for $\ket{\uparrow}$ the distribution is nonzero on only two phase points so that every trajectory starts at $s^z=+1$ with $(s^x,s^y)=(+1,+1)$ or $(-1,-1)$, each with probability $1/2$.
Fig.~\ref{fig:single-spin} shows the resulting average magnetisation $\langle\sigma^z\rangle(t)$.
Each curve represents an average over $5\times10^{4}$ trajectories, evolved with a fixed time step $\Omega_{\rm R}\,\Delta t=5\times 10^{-3}$.
On resonance both ``TWA with star product'' and ``TWA w/o star product'' are indistinguishable from the exact solution: they follow the damped Rabi oscillations and relax to the unpolarised steady state $\langle\sigma^z\rangle=0$.
At finite detuning the transient oscillations are still captured in both cases, but the steady states separate:  with the star product treated exactly, the trajectory ensemble relaxes to the value $\langle\sigma^z\rangle\approx-0.67$, whereas ``TWA w/o star product'' settles at $\langle\sigma^z\rangle\approx-0.56$, deviating from the exact steady-state magnetisation by approximately $15\%$.

\begin{table}[t]
\caption{\label{tab:cartesian-eom}%
\textbf{TWA equations of motion in Cartesian components.}
Contribution of each term of the Hamiltonian as well as the jump operators to the stochastic equations of motion.
Each dissipative contribution carries its own independent unit white noise, $\langle\xi_i(t)\,\xi_j(t')\rangle=\delta_{ij}\,\delta(t-t')$, and is written in It\^o convention.
The interaction enters through $\mathcal V_i=\sum_{j\neq i}V_{ij}\,n(s_j^z)$, with $n(s^z)=\tfrac12(1+s^z)$ the symbol of $\hat n$.
The multiplicative noise amplitudes are $B_{\downarrow \uparrow}=(\Gamma_{\downarrow \uparrow}[(1\pm s_i^z)^2+2]/[3-(s_i^z)^2])^{1/2}$ with the $+ (-)$ sign for the decay (pump) $\Gamma_{\downarrow}(\Gamma_\uparrow)$.
In the main text we only consider decay as a source of dissipation with $\Gamma_\downarrow=\Gamma$ and $\Gamma_\uparrow=\gamma=0$.}
\begin{ruledtabular}
\begin{tabular}{ll}
Term & Contribution\\
\colrule
\multicolumn{2}{l}{\emph{Coherent contributions}}\\[2pt]
\multirow{3}{*}{laser drive, $\Omega_{\mathrm R}\,\hat\sigma_i^x$}
  & $\dot s_i^x = 0$\\
  & $\dot s_i^y = -\Omega_{\mathrm R}\,s_i^z$\\
  & $\dot s_i^z = +\Omega_{\mathrm R}\,s_i^y$\\[4pt]
\colrule\noalign{\vskip 4pt}
\multirow{3}{*}{detuning, $\tfrac{\Delta}{2}\,\hat\sigma_i^z$}
  & $\dot s_i^x = -\Delta\,s_i^y$\\
  & $\dot s_i^y = +\Delta\,s_i^x$\\
  & $\dot s_i^z = 0$\\[4pt]
\colrule\noalign{\vskip 4pt}
\multirow{3}{*}{interaction, $\sum_{j\neq i}V_{ij}\,\hat n_i\hat n_j$}
  & $\dot s_i^x = -\mathcal V_i\,s_i^y$\\
  & $\dot s_i^y = +\mathcal V_i\,s_i^x$\\
  & $\dot s_i^z = 0$\\[4pt]
\colrule
\multicolumn{2}{l}{\emph{Dissipative contributions}}\\[2pt]
\multirow{3}{*}{decay, $\sqrt{\Gamma_\downarrow}\,\hat\sigma_i^-$}
  & $\dot s_i^x = -\tfrac{\Gamma_\downarrow}{2}\,s_i^x-B_\downarrow\,s_i^y\,\xi_i$\\
  & $\dot s_i^y = -\tfrac{\Gamma_\downarrow}{2}\,s_i^y+B_\downarrow\,s_i^x\,\xi_i$\\
  & $\dot s_i^z = -\Gamma_\downarrow\,(1+s_i^z)$\\[4pt]
\colrule\noalign{\vskip 4pt}
\multirow{3}{*}{pump, $\sqrt{\Gamma_\uparrow}\,\hat\sigma_i^+$}
  & $\dot s_i^x = -\tfrac{\Gamma_\uparrow}{2}\,s_i^x-B_\uparrow\,s_i^y\,\xi_i$\\
  & $\dot s_i^y = -\tfrac{\Gamma_\uparrow}{2}\,s_i^y+B_\uparrow\,s_i^x\,\xi_i$\\
  & $\dot s_i^z = +\Gamma_\uparrow\,(1-s_i^z)$\\[4pt]
\colrule\noalign{\vskip 4pt}
\multirow{3}{*}{dephasing, $\sqrt{\gamma}\,\hat\sigma_i^z$}
  & $\dot s_i^x = -2\gamma\,s_i^x-2\sqrt{\gamma}\,s_i^y\,\xi_i$\\
  & $\dot s_i^y = -2\gamma\,s_i^y+2\sqrt{\gamma}\,s_i^x\,\xi_i$\\
  & $\dot s_i^z = 0$\\
\end{tabular}
\end{ruledtabular}
\end{table}

\begin{figure*}[htb!]
    \centering
    \includegraphics[width=0.99\linewidth]{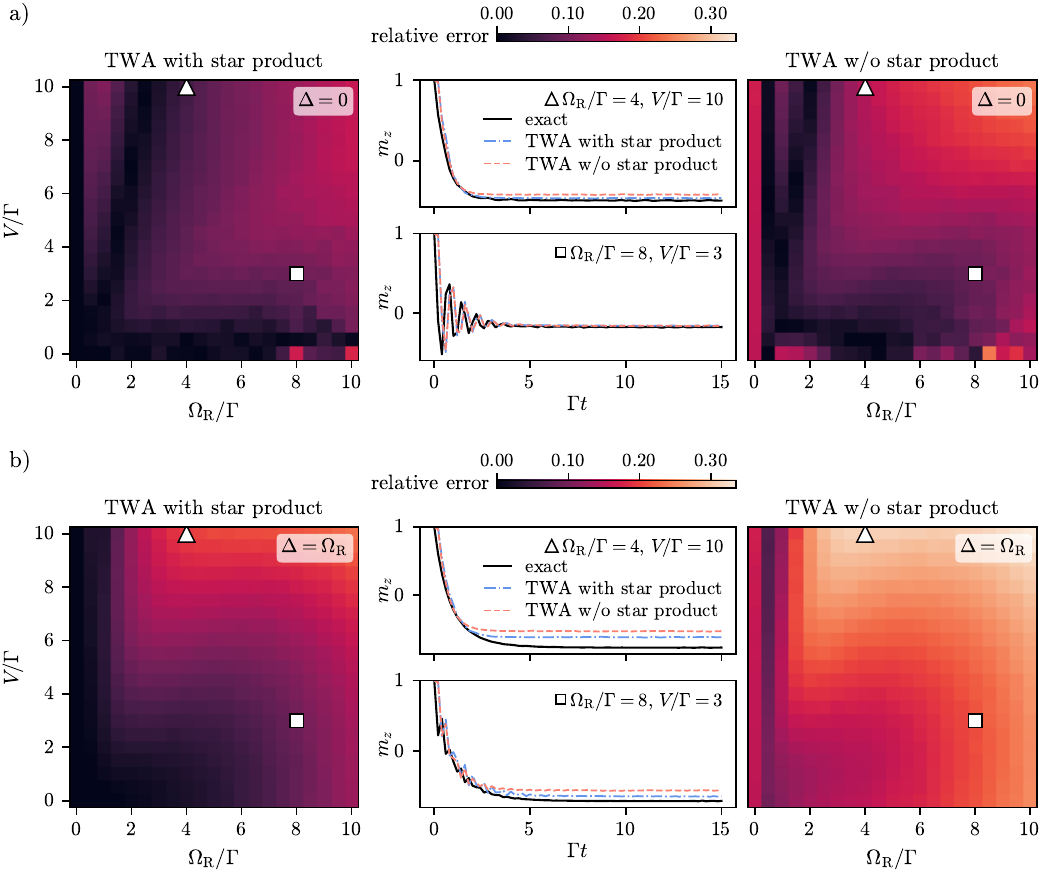}
  \caption{\textbf{Accuracy of the TWA with and without the star-product corrections.}
  The colour maps show the relative error of the steady-state, site-averaged magnetisation of an $N=10$ Rydberg chain for the TWA with the jump operators evaluated with star-product corrections (left) and without it (right) for (a)~resonant drive, $\Delta=0$ and (b)~finite detuning, $\Delta=\Omega_{\mathrm R}$.
  All the heatmaps share the same colour scale.
  The middle panels show the time evolution of $m_z$ at the parameters marked on the maps by the triangle ($\Omega_{\mathrm R}/\Gamma=4$, $V/\Gamma=10$, strongly interacting) and the square ($\Omega_{\mathrm R}/\Gamma=8$, $V/\Gamma=3$, weakly interacting), compared with the exact quantum jump Monte Carlo solution (black).}
  \label{fig:twa_phase_diagram}
\end{figure*}

We can understand why the ``TWA w/o star product'' performs well in the absence of detuning, yet deviates once the detuning is finite, as follows.
On resonance, the strong drive equalises the populations of the two levels, leaving a nearly unpolarised steady state that is less sensitive to the precise form of the dissipative terms.
A finite detuning renders the drive less effective, and the steady-state polarisation is then set by the competition between coherent excitation and decay, becoming sensitive to the dissipative terms at leading order.
Since all coherent contributions, including the detuning, enter both sets of equations identically, the detuning merely exposes an error that originates entirely in the dissipative sector, the only place where the two mappings, namely the $\mathrm{SU}(2)$ star variant compared to ordinary multiplication of jump operators differ.

As a many-body benchmark we consider the driven-dissipative chain of Eq.~\eqref{eq:ryd-ham} with van der Waals interactions $V_{ij}=V/r_{ij}^{6}$ with $r_{ij}$ denoting the distance between two sites in units of the lattice spacing. 
The relaxation dynamics of this model was recently simulated with the TWA in Ref.~\cite{noel2026quantum} and, benchmarked against exact methods, found to be accurate over a broad range of parameters.
All spins are initialised in the $\ket{\uparrow}$ state, sampled site by site from the two-point discrete Wigner distribution introduced above. 
We consider a chain with $N=10$ sites with periodic boundary conditions, and results are averaged over an ensemble of $10^4$ trajectories.
We propagate the two sets of stochastic equations compared in Fig.~\ref{fig:single-spin} for the Rydberg chain as well: the equations of Table~\ref{tab:cartesian-eom}, in which products of jump-operator symbols are evaluated with the $\mathrm{SU}(2)$ star product, and the variant in which they are replaced by ordinary multiplication using a time step $\Gamma\,\Delta t=5\times10^{-3}$ in both cases.
The reference curve was obtained from a quantum-trajectory unravelling of the Lindblad master equation with $10^3$ trajectories using the package QuantumToolbox.jl \cite{Mercurio_2025}.

As a measure for comparing the TWA with the reference we take the site-averaged magnetisation
\begin{equation}
  m_z
  =\frac{1}{N}\sum_{i=1}^{N}
   \langle\hat\sigma_{i}^{z}\rangle\,,
  \label{eq:mag}
\end{equation}
and compute its relative error to the exact steady state 
\begin{equation}
  \varepsilon
  =\frac{\bigl|\,m_{z}^{\;\mathrm{TWA}}
        -m_{z}^{\;\mathrm{exact}}\bigr|}
        {\bigl|m_{z}^{\;\mathrm{exact}}\bigr|}\,,
  \label{eq:relerr}
\end{equation}
across the drive-to-decay and interaction-to-decay ratios $\Omega_{\rm R}/\Gamma$ and $V/\Gamma$ in the range $\Omega_{\mathrm{R}}/\Gamma,\,V/\Gamma\in[0,10]$.
All simulations run to $\Gamma t=30$, and steady-state values entering Eq.~\eqref{eq:relerr} are extracted as time averages over the window $\Gamma t\in[24,30]$, well within the stationary regime.
The results are shown in Fig.~\ref{fig:twa_phase_diagram} for the resonant case (a) as well as at finite detuning (b).

In each panel the outer plots show $\varepsilon$ for the TWA with (left) and without (right) the star-product evaluation of the jump operators, while the central column displays the underlying time evolution of $m_{z}$ at two representative points, marked by a triangle ($\Omega_{\mathrm{R}}/\Gamma=4$, $V/\Gamma=10$) and a square ($\Omega_{\mathrm{R}}/\Gamma=8$, $V/\Gamma=3$).
On resonance, Fig.~\ref{fig:twa_phase_diagram}(a), the two schemes perform comparably: the relative error of either variant remains at or below the ten-percent level over most of the $(\Omega_{\mathrm{R}}/\Gamma,\,V/\Gamma)$ plane, and the time evolutions at the two marked points show both approximations following the exact relaxation dynamics. 
Only when drive and interaction are simultaneously strong does the ``TWA w/o star product'' become noticeably less accurate.
Here, errors grow towards $25\%$ in the upper right corner of the heatmap, compared to below $20\%$ for the star-product variant.
In the absence of driving, $\Omega_{\mathrm{R}}=0$ also exposes a systematic deviation: for pure decay the ``TWA w/o star product'' relaxes to $\langle\hat\sigma^{z}\rangle\approx-0.84$ instead of the value $-1$, visible as the coloured left edge of the right-hand heatmaps.
The star-product variant instead reproduces this limit exactly: with the drive absent the interaction drops out of the $\langle\hat\sigma^{z}\rangle$ dynamics, so this column isolates the dissipative sector, which the star-product mapping treats without approximation.

The picture changes at finite detuning $\Delta=\Omega_{\mathrm{R}}$, Fig.~\ref{fig:twa_phase_diagram}(b).
The ``TWA w/o star product'' develops a systematic steady-state error of $20\%$--$30\%$ across essentially the entire driven regime, largest where drive and interaction are simultaneously strong. 
Treating the star product exactly removes the dissipative contribution to this error, since decay then enters the stochastic equations without approximation.
The relative error stays below $10\%$ over most of the plane and approaches $20\%$ only in the strongly interacting corner, where the truncated interaction terms dominate the residual deviation.
The time evolution plots make the origin of this difference explicit: without the star product the magnetisation relaxes to a visibly displaced steady state, in direct analogy to the single-spin case discussed above, while the ``TWA with star product'' follows the exact relaxation closely at large drive and weak interaction (square marker, explicitly shown in the middle inset time evolution plots).
The residual error of the star-product TWA grows systematically with both $\Omega_{\mathrm{R}}/\Gamma$ and $V/\Gamma$, consistent with the build-up of genuine many-body correlations that any truncated-Wigner scheme neglects. 
The star product thus removes the systematic error in the treatment of single-spin dissipation while the remaining deviation is the expected semiclassical truncation error.

\section{Summary} 
\label{sec:summary}
We have established an $\mathrm{SU}(2)$ path-integral formulation of the TWA for dissipative spin-$1/2$ systems.
An analogous correspondence between phase-space methods and Keldysh path integrals has long been known for bosonic systems.
However, its extension to open spin systems is subtle due to the curved geometry of spin phase space and the nontrivial representation of operator products.
By consistently incorporating the curved phase-space structure, we derived stochastic equations of motion that are equivalent to those obtained from the continuous formulation of spin TWA.
In particular, our approach accurately captures decay processes and reproduces the exact dynamics of a single dissipative spin, which resolves the inconsistencies encountered in previous path-integral formulations \cite{hosseinabadi2025user, hartmann2026truncatedwignerapproximationspins}.

We have subsequently benchmarked the resulting stochastic equations against exact solutions of the corresponding master equations.
For a single driven-decaying spin, for which the star-product equations are exact, the simulations reproduce the master-equation results, whereas replacing the $\mathrm{SU}(2)$ star product by the ordinary product of symbols displaces the steady-state magnetisation considerably when the detuning is comparable to the drive.
The dissipative star-product corrections are thus required for an exact treatment of the single-spin dynamics and cannot be neglected.

For many-body interacting spins, by contrast, the method is no longer exact: the interactions generate terms coupling different sites, and their truncation, discussed in Sec.~\ref{sec:manybody}, is the defining approximation of the many-body TWA. 
We benchmarked this approximation on the driven-dissipative Rydberg chain. 
In all parameter regimes studied, the star-product TWA performs at least as well as the ordinary-product variant, and at finite detuning considerably better. 
Generally, its accuracy decreases with growing interaction strength, since the discarded terms carry a coefficient of order $V$. 
Stronger decay, on the other hand, improves it, since the steady state is then established on the timescale set by the inverse decay rate, before the effect of the discarded terms, accumulating over times of order $1/V$, becomes appreciable.

Beyond providing a field-theoretic foundation for dissipative spin TWA, the present framework offers a systematic starting point for future developments for interacting spins. 
The mixed-derivative interaction terms discarded in Sec.~\ref{sec:manybody} can be accounted for by suitable extensions of the method.
The cross term $\theta_{q,i}\,\phi_{q,j}$ becomes a genuine noise in the doubled phase space of positive-$P$-type representations \cite{drummond1980generalised, barry2008qubit, ng2011exact}, at the price of complex trajectories and their well-known sampling instabilities.
For bosons, sufficiently strong damping is known to mitigate these instabilities \cite{gilchrist1997positive}, and for driven-dissipative spin-boson networks a positive-$P$ representation built on spin coherent states has indeed been observed to be stabilised by strong drive and dissipation \cite{mandt2015stochastic}, making driven-dissipative spin lattices a promising target for such an extension.
An alternative route to recovering the discarded terms is a dissipative generalisation of the cluster truncated Wigner approximation \cite{Wurtz_2018}, which would retain the mixed derivatives within each cluster of exactly treated spins and discard only their intercluster counterparts, making the truncation systematically improvable with the cluster size.

\begin{acknowledgments}
We thank M.~Jirasek and A.~N.~Mikheev for fruitful discussions and collaboration on related projects.
This work is supported by the ERC grant OPEN-2QS (Grant No. 101164443).
I.L. acknowledges funding from the Deutsche Forschungsgemeinschaft (DFG, German Research Foundation) through the Research Unit FOR 5413/1, Grant No. 465199066 and through the Research Unit FOR 5522/1, Grant No. 499180199, as well as through JST-DFG 2024: Japanese-German Joint Call for Proposals on ``Quantum Technologies'' (Japan-JST-DFG-ASPIRE 2024) under DFG Grant No. 55456179.
M.F. acknowledges funding from the Deutsche Forschungsgemeinschaft through SFB TR 185, Project No. 277625399.
\end{acknowledgments}

\appendix 

\section{Differential identities}
\label{app:diff-id}
The dissipative sector of the main text rests on the kernel identities \eqref{eq:onesided}, in their explicit form \eqref{eq:diffident}, and on the derivative-basis expansions \eqref{eq:rec-expansion} and \eqref{eq:nj-expansion}. 
We derive them here from a single algebraic property of the Stratonovich--Weyl kernel.

\subsection{A derivative basis for the kernel}
\label{app:basis}

In the parametrisation $\hat\Delta(\Omega)=\tfrac12\bigl(\mathbb 1+\sqrt3\,\vec n\cdot\vec\sigma\bigr)$ with $\vec n=(\sin\theta\cos\phi,\sin\theta\sin\phi,\cos\theta)$, the kernel and its angular derivatives read, in the $\sigma^z$ eigenbasis,
\begin{equation}
\hat\Delta=\frac12\!\begin{pmatrix}1+\sqrt3\cos\theta & \sqrt3\sin\theta\,e^{-i\phi}\\[2pt]
\sqrt3\sin\theta\,e^{+i\phi} & 1-\sqrt3\cos\theta\end{pmatrix},
\label{eq:Amatrix}
\end{equation}
and $\partial_\theta\hat\Delta$, $\partial_\phi\hat\Delta$, $\partial_\phi^2\hat\Delta$ follow
by differentiation. 
At every $\Omega$ away from the coordinate poles $\theta=0,\pi$ (where $\partial_\phi\hat\Delta$ and $\partial_\phi^2\hat\Delta$ vanish), the four Hermitian matrices
\begin{equation}
\bigl\{\,\hat\Delta,\ \partial_\theta\hat\Delta,\ \partial_\phi\hat\Delta,\ \partial_\phi^2\hat\Delta\,\bigr\}
\label{eq:derivbasis}
\end{equation}
are linearly independent and hence span the space of $2\times2$ matrices. 
Consequently, multiplying $\hat\Delta$ on either side by a fixed operator produces a matrix that can be re-expanded in the basis \eqref{eq:derivbasis}: left or right multiplication of the kernel is represented by a differential operator acting on $\hat\Delta$.

\subsection{Kernel identities for $\hat \sigma ^{\pm}$}
\label{app:symbolmaps}
Applying this to $\hat\sigma^{-}$ from the left and to $\hat\sigma^{+}$ from the right, and fixing the coefficient functions by equating the four matrix entries, yields the identities \eqref{eq:diffident} of the main text. 
These are the $\hat\sigma^{\pm}$ analogues of the illustrative $\hat\sigma^{z}\hat\Delta$ identity of Ref.~\cite{mink2022hybrid}. 
The check is elementary, with \eqref{eq:Amatrix},
\begin{equation}
\begin{aligned}
\hat\sigma^{-}\hat\Delta&=\frac12\!\begin{pmatrix}0 & 0\\[2pt]
1+\sqrt3\cos\theta & \sqrt3\sin\theta\,e^{-i\phi}\end{pmatrix},\\[4pt]
\hat\Delta\,\hat\sigma^{+}&=\frac12\!\begin{pmatrix}0 & 1+\sqrt3\cos\theta\\[2pt]
0 & \sqrt3\sin\theta\,e^{+i\phi}\end{pmatrix},
\end{aligned}
\end{equation}
and one verifies entry by entry that the right-hand sides of \eqref{eq:diffident} reproduce these matrices. 
In fact only the first identity needs to be computed: since the kernel and its derivatives are Hermitian, the second is the Hermitian conjugate of the first, $\hat\Delta\,\hat\sigma^{+}=(\hat\sigma^{-}\hat\Delta)^{\dagger}$, so its coefficient functions are the complex conjugates.

\subsection{The jump term}
\label{app:recycling}

In the main text the jump entry of the decay channel is evaluated by
collapsing
$\hat\sigma^{-}\hat\Delta\,\hat\sigma^{+}
=\operatorname{tr}[\hat\sigma^{+}\hat\sigma^{-}\hat\Delta]\,
\hat\sigma^{-}\hat\sigma^{+}
=\tfrac12(1+\sqrt3\cos\theta)\,\hat\sigma^{-}\hat\sigma^{+}$,
so the only ingredient of \eqref{eq:rec-expansion} not derived there is the expansion of the projector $\hat\sigma^{-}\hat\sigma^{+}$ in the basis \eqref{eq:derivbasis}. 
Writing $\hat\sigma^{-}\hat\sigma^{+} =a\,\hat\Delta+b\,\partial_\theta\hat\Delta+c\,\partial_\phi\hat\Delta +d\,\partial_\phi^2\hat\Delta$ and matching the four matrix entries with \eqref{eq:Amatrix} gives
\begin{equation}
a=1,\quad
b=\frac{\sqrt3\cos\theta+1}{\sqrt3\,\sin\theta},\quad
c=0,\quad
d=\frac{\sqrt3+\cos\theta}{\sqrt3\,\sin^2\theta}.
\label{eq:proj-expansion}
\end{equation}
Two of the coefficients can be read off directly: the trace fixes $a=\operatorname{tr}[\hat\sigma^{-}\hat\sigma^{+}]=1$, since $\operatorname{tr}\hat\Delta=1$ while the derivatives of the kernel are traceless, and $c=0$ because $\partial_\phi\hat\Delta$ carries the only imaginary part of the phase-stripped off-diagonal entry, which must vanish separately. 
Multiplying \eqref{eq:proj-expansion} by the prefactor $\tfrac12(1+\sqrt3\cos\theta)$ reproduces \eqref{eq:rec-expansion}; the same entry-matching applied to $\tfrac12\{\hat\sigma^{+}\hat\sigma^{-},\hat\Delta\}$ yields \eqref{eq:nj-expansion}.

\section{Replacing the star product with the ordinary multiplicative product in the dissipative part of the action}

In this section, we compute the dissipative contributions to the action by replacing the star product of symbols with ordinary multiplication.

\subsection{Decay}
\label{sec:app-multi}
The decay contributions to the dissipative action have been evaluated via the adequate operator mapping in the main text in Sec.~\ref{sec:dissipative}.
Here, we show instead what one would get by replacing the $\mathrm{SU}(2)$ star product with ordinary multiplication of phase-space symbols.
The contributions to the ordinary product come from
\begin{equation}
L_+ L^*_- = \tfrac{ 3\Gamma} {4}\,\sin\theta_+\sin\theta_-\,e^{-i\phi_q}
\end{equation}
and
\begin{equation}
\tfrac12 L^*_\pm L_\pm=\tfrac{3\Gamma}{8}\sin^2\theta_\pm,
\end{equation}
where the relative phase $\phi_q=\phi_+-\phi_-$ survives only in the cross-branch term, the same-branch terms being real. 
Inserting the inverse of \eqref{eq:pairing} and expanding each sine,
\begin{equation}
\begin{aligned}
\sin\theta_\pm
&=\sin\theta_{\rm cl}\cos\tfrac{\theta_q}{2}\pm\cos\theta_{\rm cl}\sin\tfrac{\theta_q}{2}\\
&=\sin\theta_{\rm cl}\Bigl(1-\tfrac{\theta_q^2}{8}\Bigr)
\pm\tfrac{\theta_q}{2}\cos\theta_{\rm cl}
+\mathcal O(\theta_q^3),
\end{aligned}
\end{equation}
the cross- and same-branch factors become
\begin{equation}
\sin\theta_+\sin\theta_-=\sin^2\theta_{\rm cl}-\tfrac{\theta_q^2}{4},
\end{equation}
\begin{equation}
\tfrac12(\sin^2\theta_++\sin^2\theta_-)=\sin^2\theta_{\rm cl}+\tfrac{\theta_q^2}{4}\cos2\theta_{\rm cl}.
\end{equation}
Multiplying the jump term by $e^{-i\phi_q}=1-i\phi_q-\tfrac12\phi_q^2+\dots$
and subtracting the no-jump sum, the dissipative contribution becomes
\begin{equation}
\begin{aligned}
&L_+L^*_--\tfrac12 L^*_+L_+-\tfrac12 L^*_-L_-\\
&\quad=\frac{3\Gamma}{4}\Bigl[-i\phi_q\sin^2\theta_{\rm cl}-\tfrac12\phi_q^2\sin^2\theta_{\rm cl}
-\tfrac{\theta_q^2}{2}\cos^2\theta_{\rm cl}\Bigr],
\end{aligned}
\end{equation}
using $\tfrac14(1+\cos2\theta_{\rm cl})=\tfrac12\cos^2\theta_{\rm cl}$. The overall
$-i$ turns the $\phi_q$ term real (drift) and the $q^2$ terms imaginary (noise),
\begin{equation}
\begin{aligned}
S_L^{(\text{prod})}
\simeq{}&-\tfrac{3\Gamma}{4}\!\int dt\,\phi_q\sin^2\theta_{\rm cl}\\
&+\tfrac{3i\Gamma}{8}\!\int dt\,\bigl[\phi_q^2\sin^2\theta_{\rm cl}
+\theta_q^2\cos^2\theta_{\rm cl}\bigr].
\end{aligned}
\label{eq:SL-prod}
\end{equation}
The two results \eqref{eq:SL-prod} and \eqref{eq:SL-star} differ in both sectors.
The drift $-\tfrac{3\Gamma}{4}\sin^2\theta_{\rm cl}$ of the ordinary product is replaced by $-\tfrac{\Gamma}{2}(1+\sqrt3\cos\theta_{\rm cl})$, which is what pulls $\theta_{\rm cl}$ towards the decay fixed point $\cos\theta_{\rm cl}=-1/\sqrt3$, i.e.\ $\langle\sigma^z\rangle=-1$.
The ordinary-product drift instead relaxes every trajectory to the south pole of the radius-$\sqrt3$ sphere.
More strikingly, the star result \eqref{eq:SL-star} has no $\phi_q^2$ term at all. 
The $\phi_q^2\sin^2\theta_{\rm cl}$ noise of the ordinary product \eqref{eq:SL-prod} is therefore a spurious noise channel absent from the divergence form \eqref{eq:fp}. 

\subsection{Dephasing}
\label{app:multi-deph}

We now repeat the comparison for the dephasing channel, $\hat L=\sqrt{\gamma}\,\hat\sigma^z$, and show that here the ordinary multiplicative product reproduces the star-product result exactly at the retained order.

We first record the exact evaluation, following Sec.~\ref{sec:dissipative}.
The jump operator is Hermitian and squares to the identity, $\hat L^\dagger\hat L=\gamma\,\mathbb 1$, so the no-jump entries are trivial and the dissipator collapses to a double commutator,
\begin{equation}
\mathcal D_\gamma\hat\rho
=\gamma\bigl(\hat\sigma^z\hat\rho\,\hat\sigma^z-\hat\rho\bigr)
=-\tfrac{\gamma}{2}\,\bigl[\hat\sigma^z,\bigl[\hat\sigma^z,\hat\rho\bigr]\bigr].
\label{eq:deph-dcomm}
\end{equation}
On the kernel, the commutator is evaluated by the exact first-order identity \eqref{eq:spinbracket} with the symbol $\sqrt3\cos\theta$ of $\hat\sigma^z$ from \eqref{eq:pauli-symbols},
\begin{equation}
[\hat\sigma^z,\hat\Delta]=2i\,\partial_\phi\hat\Delta .
\label{eq:sz-comm}
\end{equation}
Since the differential operator in \eqref{eq:sz-comm} carries a constant coefficient, it commutes with multiplication by the fixed matrix $\hat\sigma^z$, and the outer commutator simply iterates \eqref{eq:sz-comm},
\begin{equation}
\mathcal D_\gamma\hat\Delta
=-\tfrac{\gamma}{2}\,\bigl(2i\,\partial_\phi\bigr)^{2}\hat\Delta
=2\gamma\,\partial_\phi^{2}\hat\Delta .
\label{eq:kernel-deph}
\end{equation}

The jump entry can also be checked directly, as done for decay in \eqref{eq:rec-expansion}: since $\hat\sigma^z$ anticommutes with $\hat\sigma^{x,y}$, conjugation flips the transverse components of the Bloch vector,
\begin{equation}
\begin{aligned}
\hat\sigma^z\hat\Delta(\theta,\phi)\,\hat\sigma^z
&=\tfrac12\bigl(\mathbb 1+\sqrt3\,(n_z\hat\sigma^z-n_x\hat\sigma^x
-n_y\hat\sigma^y)\bigr)\\
&=\hat\Delta(\theta,\phi+\pi),
\end{aligned}
\label{eq:sz-sandwich}
\end{equation}
and matching the matrix entries against the basis \eqref{eq:der-basis} as in App.~\ref{app:recycling} gives
\begin{equation}
\hat\sigma^z\hat\Delta\,\hat\sigma^z=\bigl(1+2\,\partial_\phi^{2}\bigr)\hat\Delta ,
\label{eq:sz-sandwich-basis}
\end{equation}
with the coefficient of $\hat\Delta$ again fixed to one by the trace, $\operatorname{tr}[\hat\sigma^z\hat\Delta\hat\sigma^z]=\operatorname{tr}\hat\Delta=1$.
Subtracting the no-jump entry $\tfrac12\{\hat L^\dagger\hat L,\hat\Delta\}=\gamma\hat\Delta$ reproduces \eqref{eq:kernel-deph}.

Equation \eqref{eq:kernel-deph} is of the form \eqref{eq:kernel-diss} with $A_\theta=0$ and $B_{\phi\phi}=4\gamma$: dephasing generates no drift and adds the pure azimuthal diffusion $2\gamma\,\partial_\phi^2\chi$ to the Fokker--Planck equation \eqref{eq:fp}, in line with its physical reading as a random rotation about the $z$ axis.
Under the substitution \eqref{eq:prepoint} the dissipative symbol reads
\begin{equation}
\ell_D^{(\gamma)}=2\gamma\,(i\tilde\phi)^{2},
\label{eq:ellD-deph}
\end{equation}
exact for the same reason as \eqref{eq:ellD-star}, and with the change of variables \eqref{eq:clq-def}, i.e.\ $\tilde\phi=-\tfrac{\sqrt3}{2}\sin\theta_{\rm cl}\,\theta_q$, the contribution to the action, $S_\gamma=-i\!\int\!dt\,\ell_D^{(\gamma)}$, becomes
\begin{equation}
S_\gamma^{(\star)}
=\frac{3i\gamma}{2}\!\int\!dt\,\sin^{2}\theta_{\rm cl}\,\theta_q^{2}.
\label{eq:SL-deph-star}
\end{equation}
The action is purely quadratic in the quantum fields: dephasing contributes no drift, and only adds the entry $D_\gamma=3\gamma\sin^{2}\theta_{\rm cl}$ to the noise kernel of \eqref{eq:EED}.
Note also that neither $\phi_q$ nor $\tilde\theta$ appears, so by the crosswise pairing of \eqref{eq:EED} the equation of $\theta_{\rm cl}$ receives neither drift nor noise.
Repeating the steps of Sec.~\ref{sec:eom}, the noise enters the equation of motion of $\phi_{\rm cl}$ with amplitude $(2/\sqrt3\sin\theta_{\rm cl})\sqrt{D_\gamma}=2\sqrt\gamma$,
\begin{equation}
\dot\theta_{\rm cl}=\ldots,\qquad
\dot\phi_{\rm cl}=\ldots+2\sqrt{\gamma}\,\xi(t),
\label{eq:eom-deph}
\end{equation}
with unit white noise $\langle\xi(t)\xi(t')\rangle=\delta(t-t')$, the dots denoting the contributions of the other terms. 
Rewriting this stochastic rotation about the $z$ axis in Cartesian components in the It\^o convention yields the dephasing row of Table~\ref{tab:cartesian-eom}.

We now repeat the evaluation with the ordinary multiplicative product of the branch symbols. 
The symbol of the jump operator follows from
\eqref{eq:pauli-symbols},
\begin{equation}
L(\theta,\phi)=\sqrt{3\gamma}\,\cos\theta ,
\label{eq:Ldeph-symbol}
\end{equation}
and is real and independent of $\phi$: the branch symbols $L_\pm=L^*_\pm=\sqrt{3\gamma}\cos\theta_\pm$ carry no relative phase, in contrast to the factor $e^{-i\phi_q}$ of the decay channel.
The dissipative combination therefore assembles into a perfect square,
\begin{equation}
\begin{aligned}
L_+L^*_--\tfrac12 L^*_+L_+-\tfrac12 L^*_-L_-
&=-\tfrac12\bigl(L_+-L_-\bigr)^{2}\\
&=-\tfrac{3\gamma}{2}\bigl(\cos\theta_+-\cos\theta_-\bigr)^{2}.
\end{aligned}
\label{eq:deph-square}
\end{equation}
Inserting the pairing \eqref{eq:pairing},
\begin{equation}
\cos\theta_+-\cos\theta_-
=-2\sin\theta_{\rm cl}\,\sin\tfrac{\theta_q}{2}
=-\sin\theta_{\rm cl}\,\theta_q+\mathcal O(\theta_q^{3}),
\label{eq:deph-diff}
\end{equation}
so that
\begin{equation}
\ell_D^{(\gamma,\text{prod})}
=-6\gamma\,\sin^{2}\theta_{\rm cl}\,\sin^{2}\tfrac{\theta_q}{2}
=-\tfrac{3\gamma}{2}\,\sin^{2}\theta_{\rm cl}\,\theta_q^{2}
+\mathcal O(\theta_q^{4}),
\end{equation}
and the action becomes
\begin{equation}
S_\gamma^{(\text{prod})}
=\frac{3i\gamma}{2}\!\int\!dt\,\sin^{2}\theta_{\rm cl}\,\theta_q^{2}
+\mathcal O(\theta_q^{4})
=S_\gamma^{(\star)}+\mathcal O(\theta_q^{4}).
\label{eq:SL-deph-prod}
\end{equation}
The two evaluations differ only at quartic order in the quantum fields, which the semiclassical truncation discards in any case; and since the exact star result \eqref{eq:SL-deph-star} is already purely quadratic, the ordinary product reproduces the exact dephasing action, with the same vanishing drift
and the same noise, and hence the stochastic equation and trajectories.

The agreement is not accidental and rests on two properties of this dissipative channel.
First, the jump operator is Hermitian: the dissipator is then the double commutator \eqref{eq:deph-dcomm}, and the symmetric (anticommutator) halves of the identities, which carry the derivative corrections responsible for the decay-channel discrepancy, drop out entirely.
What remains is the square of a commutator, and for spin-$1/2$ the sided difference represented by a commutator is exactly linear in the quantum fields (cf.~\eqref{eq:sided-split-q}). 
The ordinary product, whose error in the branch difference \eqref{eq:deph-diff} is of order $\theta_q^{3}$, fails only at order $\theta_q^{4}$ once squared.
Second, the symbol \eqref{eq:Ldeph-symbol} depends on $\theta$ only, so the first-order operator in \eqref{eq:sz-comm} has a constant coefficient and its square \eqref{eq:kernel-deph} contains no lower-order (drift) terms. 
For a Hermitian jump operator without this axial symmetry the square of the commutator would still generate a drift from the derivatives of its own coefficients, which the ordinary product would again miss.

\bibliography{apssamp}

\end{document}